\pdfoutput=1
\newcommand*{\ATLASLATEXPATH}{./}
\documentclass[cernpreprint,texlive=2016,txfonts,UKenglish]{atlasdoc} 
\usepackage[subfigure=true]{\ATLASLATEXPATH atlaspackage}
\usepackage{\ATLASLATEXPATH atlasbiblatex}

\usepackage{\ATLASLATEXPATH atlascontribute}

\usepackage{\ATLASLATEXPATH atlasphysics}

\makeatletter
\def\blx@maxline{77}
\makeatother

\graphicspath{{logos/}{figures/}}

\RequirePackage{tikz}
\usepackage{graphicx}
\usepackage{xspace}
\usepackage{longtable}
\usepackage{multicol}
\usepackage{multirow}
\usepackage{amsmath,amssymb}
\usepackage{relsize}
\usepackage{bm}
\usepackage{url}
\usepackage{bigstrut}
\usepackage{placeins}
\usepackage{setspace}
\usepackage{hyperref}
\usepackage{booktabs}
\usepackage{amssymb}
\usepackage{changepage}
\usepackage{xcolor}

\def\evtgen{\textsc{EvtGen}\xspace}
\def\geantfour{\textsc{Geant4}\xspace}
\def\atlfasttwo{\textsc{AtlFastII}\xspace}

\def\sherpa{\textsc{Sherpa}\,2.1.1\xspace}
\def\amcnlo{\textsc{MadGraph}}
\def\mg{\textsc{MadGraph}\xspace}

\DeclareRobustCommand{\pythiaeight}{\textsc{Pythia\,8.186}\xspace}

\def\powheg{\textsc{Powheg-Box}\xspace}

\newcommand{\mBH}{\ensuremath{m_{H}}\xspace}
\newcommand{\mchi}{\ensuremath{m_{\chi}}\xspace}
\def\kt{\ensuremath{k_t}\xspace}
\def\kT{\kt}
\def\etmiss{\ensuremath{E_\mathrm{T}^{\mathrm{miss}}}\xspace}
\def\setmiss{\ensuremath{S_{E_\mathrm{T}^{\mathrm{miss}}}}\xspace}
\def\pThard{\ensuremath{\pT^{\mathrm{hard}}}\xspace}
\def\pTgg{\ensuremath{\pT^{\gamma\gamma}}\xspace}
\def\mgg{\ensuremath{m_{\gamma\gamma}}\xspace}

\usepackage{dcolumn}
\newcolumntype{.}{D{.}{.}{-1}}

\AtlasTitle{Search for dark matter in association with a Higgs boson decaying to two photons at $\sqrt{s}$ = 13 TeV with the ATLAS detector}

\author{The ATLAS Collaboration}

\AtlasRefCode{HIGG-2016-18}
\PreprintIdNumber{CERN-EP-2017-079}

\AtlasJournalRef{Phys.\ Rev.\ D 96 (2017) 112004}

\AtlasDOI{10.1103/PhysRevD.96.112004}

\AtlasAbstract{
A search for dark matter in association with a Higgs boson decaying to two photons is presented.
This study is based on data collected with the ATLAS detector, corresponding to an integrated luminosity of $36.1~\mathrm{fb}^{-1}$ of proton--proton
collisions at the LHC at a center-of-mass energy of 13 TeV in 2015 and 2016.
No significant excess over the expected background is observed.
Upper limits at 95\% confidence level are set on
the visible cross section for beyond the Standard Model physics processes, and the production cross section times branching fraction
of the Standard Model Higgs boson decaying into two photons in association with missing
transverse momentum in three different benchmark models.
Limits at 95\% confidence level are also set on the observed signal in
two-dimensional mass planes.
Additionally, the results are interpreted in terms of 90\% confidence-level limits on
the dark-matter--nucleon scattering cross section, as a function of the dark-matter particle mass, for a spin-independent scenario.
}

\begin{document}

\maketitle

\clearpage
\section{Introduction}
\label{sec:intro}
The discovery of a particle consistent with the Standard Model (SM) Higgs boson in 2012 by the ATLAS~\cite{Aad:2012tfa} and CMS~\cite{Chatrchyan:2012xdj}
collaborations has opened up new possibilities in searches for physics beyond the SM (BSM).
Although strong astrophysical evidence~\cite{Olive:2016xmw,Bertone:2004pz} indicates the possible existence of dark matter (DM),
there is no evidence yet for non-gravitational interactions between DM and SM particles.
The interaction probability of DM particles, which are produced in SM particle collisions, with a detector is expected to be tiny.
Thus, many searches for DM at the Large Hadron Collider (LHC) involve missing transverse momentum
($\etmiss$) produced in association with detectable particles ($X$+\met final states).
In other $X$+\met searches in proton--proton ($pp$) collisions, $X$ may represent a jet
or a $\gamma$/$W$/$Z$ boson, which can be emitted directly from a light quark or gluon as
initial-state radiation through SM gauge interactions. However, SM
Higgs boson radiation from initial-state partons is highly suppressed, so events
with a final state compatible with the production of a SM Higgs boson in
association with \etmiss can be sensitive probes of the structure of the BSM
physics responsible for producing DM. 
The SM Higgs boson is expected to be produced from a new interaction between DM and the SM particles~\cite{monoh,Petrov:2013nia}.
Both the ATLAS and CMS collaborations have previously searched for such topologies using
$20.3~\mathrm{fb}^{-1}$ of $pp$ collision data at $\sqrt{s} =8$~\TeV~\cite{Aad:2015yga,Aad:2015dva}, and
$2.3$--$36.1~\mathrm{fb}^{-1}$ of $pp$ collision data at $\sqrt{s}
=13$~\TeV~\cite{Aaboud:2016obm,Sirunyan:2017hnk,CERN-EP-2017-117}, considering the SM Higgs boson decay into a
pair of photons or $b$-quarks in events with missing transverse momentum.
Although the branching fraction of the SM Higgs boson decaying into a pair of photons is small, the diphoton system presented 
in this paper falls in a narrower mass range around the Higgs boson mass than in Ref.~\cite{CERN-EP-2017-117}. 
With the diphoton trigger, this channel is more sensitive in the low \met region than the channel with the SM Higgs boson decaying into a pair of $b$-quarks, which relies on the high \met trigger. 
This paper presents an updated search for DM particles ($\chi$) associated with the SM Higgs
boson ($h$) decay to a pair of photons using $36.1~\mathrm{fb}^{-1}$ of $pp$
collision data collected at $\sqrt{s}=13~\TeV$ during 2015 and 2016, where both
the integrated luminosity and the center-of-mass energy are significantly higher
than in the previously published ATLAS analysis~\cite{Aad:2015yga}.

\begin{figure}[!htp]
    \centering
    \begin{tabular}{cccc}
    \subfigure[]{\includegraphics[width=0.33\textwidth]{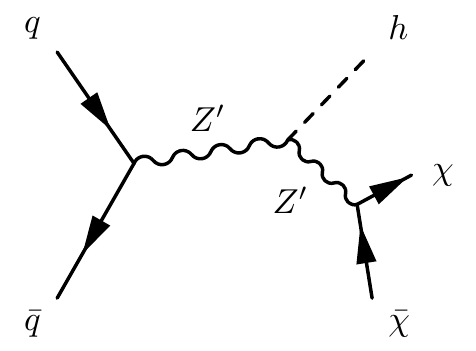}}
    \subfigure[]{\includegraphics[width=0.33\textwidth]{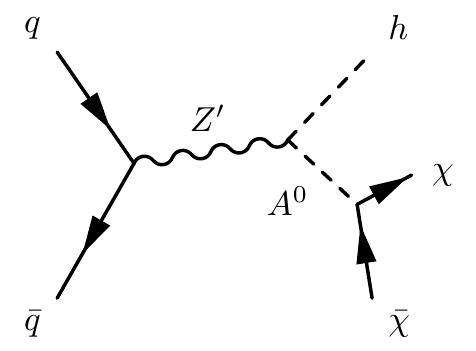}}
    \subfigure[]{\includegraphics[width=0.33\textwidth]{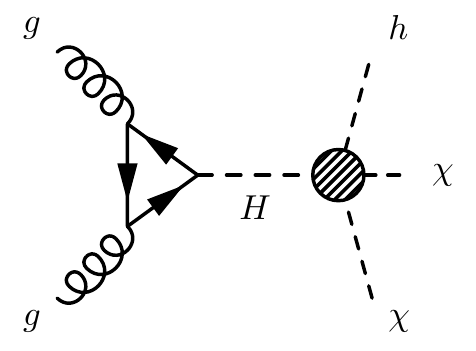}}
    \end{tabular}
      \caption{The Feynman diagrams for the production of DM ($\chi$) in association with a
      SM Higgs boson ($h$) arising from three theoretical models
      considered in this paper: (a) $Z'_{B}$ model, (b) $Z'$-2HDM
      model, (c) heavy-scalar model.}
    \label{fig:feyn}
  \end{figure}

Three benchmark models are considered in this analysis.
The leading-order (LO) Feynman diagrams representing the production
of $h$ plus $\etmiss$ in two simplified models~\cite{lhcdmf} are shown in
Figs.~\ref{fig:feyn}(a) and~\ref{fig:feyn}(b).
In the first model, a massive vector mediator $Z'$ emits a Higgs boson and subsequently decays to
a pair of Dirac fermionic DM candidates. A vector-boson mediator arises in many BSM theories through a minimal extension
to the gauge sector of the SM.
In scenarios where the DM couples to the SM only via the $Z'$ boson (i.e., the $Z'_{B}$
model~\cite{monoh} represented in Fig.~\ref{fig:feyn}(a)), the associated $U'(1)$ symmetry
ensures the stability of the DM particle.
The baryon number $B$ is associated with the gauge symmetry of $U(1)_B$, and an additional
scalar particle (referred to as a baryonic Higgs
boson) is introduced to break this symmetry spontaneously and generate the $Z'$
boson mass (denoted by $m_{Z'_B}$).
The second model (from a $Z'$-two-Higgs doublet model ($Z'$-2HDM)~\cite{z'-2hdm},
Fig.~\ref{fig:feyn}(b)) involves the $Z'$ boson decaying to the SM Higgs boson
and an intermediate heavy pseudoscalar boson $A^0$, which then decays to a pair of Dirac fermionic DM particles.
The minimum decay widths of the mediators are assumed for both the $Z'_{B}$ and $Z'\text{-2HDM}$ models
to be the sum of the partial widths for all decays into DM and quarks that are kinematically accessible~\cite{lhcdmf}.
Alongside those simplified models recommended in Ref.~\cite{lhcdmf}, a third
model (referred to as the heavy-scalar model~\cite{vonBuddenbrock:2016rmr}, Fig.~\ref{fig:feyn}(c))
introduces a heavy scalar boson $H$
produced primarily via gluon--gluon fusion (ggF)
with a mass in the range $2m_{h} < \mBH < 2m_{\mathrm{top}}$, in which
$m_{h}$ and $m_{\mathrm{top}}$ represent the masses of the SM Higgs
boson and top quark, respectively.
The upper bound on \mBH is introduced to avoid a large branching fraction for $H \to \ttbar$,
which would saturate the entire $H$ width leading to a $H\to h\chi\chi$
branching fraction close to zero. The lower bound on \mBH is required to be more than twice of $m_h$ to
ensure the SM Higgs boson is produced on-shell.
An effective quartic coupling between $h$, $H$, and $\chi$ is
considered, where the DM $\chi$ is assumed to be a scalar particle.
The decay branching fraction of $H$ to $h$ and two
$\chi$ particles is assumed to be 100\% for this model, to simplify the interpretations.
The DM mass (\mchi) is taken to be roughly half
of the SM Higgs-boson mass to ensure on-shell decay of $H\to h\chi\chi$, and to
suppress invisible decay modes of $h$, as described in Ref.~\cite{Guo:2010hq}.
While no assumptions are made here as to the nature of $H$, it can be viewed as a
part of a 2HDM+$\chi$ scenario where $H$ may be considered as the CP-even heavy scalar boson~\cite{vonBuddenbrock:2016rmr}.

This paper is organized as follows.
Section~\ref{sec:detector} gives a brief description of the ATLAS detector.
Section~\ref{sec:simulation} describes the data set and the signal and background Monte Carlo (MC) simulation samples used.
Section~\ref{sec:reconstruction} explains the event reconstruction, while
Section~\ref{sec:evtselection} outlines the optimization of the event selection and categorization.
Section~\ref{sec:backgroundest} summarizes the signal and background modeling.
Section~\ref{sec:effsys} discusses the experimental and theoretical systematic uncertainties that affect the results.
Section~\ref{sec:result} presents the results and their interpretations, and finally
a summary of the results is given in
Section~\ref{sec:summary}.
 \section{ATLAS detector}
\label{sec:detector}

The ATLAS detector~\cite{Aad:2008zzm,ATLAS-TDR-19} is
a multipurpose particle physics detector with approximately forward-backward symmetric
cylindrical geometry.\footnote{ATLAS uses a right-handed coordinate
system with its origin at the nominal interaction point (IP) in the
center of the detector and the $z$-axis along the beam pipe. The
$x$-axis points from the IP to the center of the LHC ring, and the
$y$-axis points upward. Cylindrical coordinates $(r,\phi)$ are used
in the transverse plane, $\phi$ being the azimuthal angle around the
beam pipe. The pseudorapidity is defined in terms of the polar angle
$\theta$ as $\eta=-\ln[\tan(\theta/2)]$. The photon transverse energy is $\et = E/
\cosh(\eta)$, where $E$ is its energy.}
The inner detector (ID) tracking system covers $\abseta$ < 2.5 and consists of a
 silicon pixel detector, a silicon microstrip detector and a transition
 radiation tracker (TRT).
The ID allows a precise
reconstruction of charged-particle trajectories and of decay vertices
 of long-lived particles. The ID is surrounded by a thin
superconducting solenoid providing a $2\:\mathrm{T}$ axial magnetic field.
A high-granularity lead/liquid-argon (LAr) sampling
calorimeter measures the energy and the position of
electromagnetic showers in the central ($\left| \eta \right|<1.475$) and endcap ($1.375 < \left| \eta \right| < 3.2$) regions.
It includes a presampler (for $|\eta|<1.8$) and three sampling layers up to $|\eta|<2.5$. The longitudinal and lateral segmentation of the calorimeter allows a measurement of the shower direction without assuming that the photon originates from a specific point along the beamline.
LAr sampling calorimeters with copper and tungsten absorbers are also used to measure hadronic showers in
the endcap ($1.5<\left| \eta \right|<3.2$) and forward ($3.1<\left|
\eta \right|<4.9$) regions, while a steel/scintillator-tile
calorimeter measures hadronic showers in the central
region ($\left| \eta \right|<1.7$).
The muon spectrometer surrounds the calorimeters and consists of three
large superconducting air-core toroid magnets, each with eight coils,
a system of precision tracking chambers ($\left| \eta \right|<2.7$),
and fast tracking chambers for triggering ($|\eta|<2.4$).
Reconstructed events are selected by a two-level trigger system. The first-level trigger is
hardware-based, while the second-level trigger is implemented in software~\cite{Aaboud:2016leb}.

 \section{Data and simulation samples}
\label{sec:simulation}
The analysis uses $pp$ collision data with a bunch crossing interval of $25$ ns,
collected in 2015 and 2016 at $\sqrt{s}= 13$~TeV. Only events that were recorded
in stable-beam conditions, when relevant detector components were functioning
properly, are considered.
Events were collected using a diphoton trigger
requiring two reconstructed photon candidates with transverse energies (\ET) of at least $35$ and $25$ \GeV\ for the $\et$-ordered leading and
subleading photons, respectively. The trigger efficiency with respect to the
offline-reconstructed photons, measured using the same method as described in
Ref.~\cite{Aaboud:2016tru}, is 99\%.
The data sample corresponds to an integrated luminosity of $36.1~\mathrm{fb}^{-1}$.
There are on average 25 interactions in the same bunch crossing (in-time pileup)
in this data sample.

The MC simulation of signal and background processes is used  to optimize the selection criteria, estimate uncertainties, and study the shapes
of the signal and background diphoton invariant mass (\mgg) distributions. Signal
events from $Z'_{B}$, $Z'$-2HDM, and the heavy-scalar models are generated using
\amcnlo\_aMC@NLO 2.2.3~\cite{Alwall:2014hca} at LO in quantum chromodynamics (QCD)
using the NNPDF3.0LO~\cite{Ball:2014uwa} parton distribution function (PDF) set.
Parton showering and hadronization are handled by the \mbox{\pythiaeight~\cite{Sjostrand:2007gs}}
event generator with the A14~\cite{ATL-PHYS-PUB-2014-021} set of tuned
parameters (tune), using the NNPDF2.3LO PDF set~\cite{Ball:2012cx}.
MC samples for the $Z'_{B}$ model are generated assuming a DM mass $m_\chi$
between $1$ and $1000$~\GeV\ and a range of mediator masses $m_{Z'_{B}}$ between $1$ and $2000$~\GeV.
The upper bounds of the DM and mediator masses are motivated by the limited sensitivity with the current integrated luminosity,
though in principle even heavier signals can be probed with this analysis.
The values of the coupling constants and mixing parameter are chosen following the
recommendations of the LHC Dark Matter Forum~\cite{lhcdmf}.
The couplings of the $Z'$ boson to DM particles ($g_{\chi}$), to quarks ($g_{q}$)
and to the SM Higgs boson ($g_{hZ'Z'}/m_{Z'}$) are set to 1.0, 1/3, and 1.0, respectively.
The mixing angle between the baryonic Higgs boson and the SM Higgs boson is set
to $\sin\theta = 0.3$.
The kinematic distributions predicted by the model are not strongly dependent on
the coupling parameters after similar cuts either at the particle level or at the reconstruction level,
and thus all samples are generated using this set of parameters.
Similarly, the MC samples for the $Z'$-2HDM model are generated for ranges of values of the
mediator mass $m_{Z'}=400$ to $1400$ GeV\ and pseudoscalar boson mass
$m_{A^0}=200$ to $450$ GeV\ for which the search is sensitive.
The masses of the neutral CP-even scalar ($H^0$) and the charged scalars ($H^{\pm}$) from $Z'$-2HDM model are set to 300 GeV.
The DM mass $m_\chi$ is set to $100$ GeV, the ratio of the two-Higgs-doublet
vacuum expectation values is set to $\tan \beta = 1.0$ and the coupling constant
between the $Z'$, Higgs, and pseudoscalar bosons is set to $g_{Z'} = 0.8$,
as suggested by Ref.~\cite{lhcdmf}.
The kinematics do not change appreciably when different values of $\tan\beta$ and $g_{Z'}$ are used.
Moreover, in the signal process, as the DM pairs are produced as a result of the $A^0$ decay,
there are minimal kinematic changes when the $A^0$ is produced on-shell.
For the heavy-scalar model, the events are generated with \mBH in steps of 10~\GeV\ in the intervals
$260~\le\mBH\le 270~\GeV$ and $300~\le\mBH\le 350~\GeV$ and in steps of 5 \GeV\ in the intervals $270~\le\mBH\le 300~\GeV$ for
$\mchi = 60$~\GeV. This mass value ($\mchi = 60$~\GeV) was chosen in order to 
ensure on-shell decay of $H\to h\chi\chi$, and at the same time to suppress invisible decay modes of $h$. 
The impact of this choice on the expected sensitivity is negligible for $\chi$ mass from 50~\GeV\ to 70~\GeV. 
The mass point $\mBH = 275$~\GeV\ and $\mchi = 60$~\GeV\ is taken as an example of
the kinematics that depend on the value of \mBH in this model.

The dominant backgrounds are resonant SM $h \rightarrow \gamma\gamma$, and non-resonant
$\gamma\gamma$, $\gamma$+jet, $W\gamma$, $Z\gamma$, $W\gamma\gamma$, and $Z\gamma\gamma$ production.
For the resonant SM Higgs-boson production, events from $Wh$ and $Zh$ processes
are generated by \pythiaeight with the A14 tune and the NNPDF2.3LO PDF set.
The ggF and vector-boson fusion (VBF) samples are generated by \powheg 2~\cite{Nason:2004rx,Alioli:2008tz,Nason:2009ai,Alioli:2010xd}
interfaced to \pythiaeight with
the AZNLO~\cite{ATL-PHYS-PUB-2013-017} tune and the CT10 PDF
set~\cite{Lai:2010vv}. Samples of $t\overline{t}h$ events are generated with
\mbox{\amcnlo\_aMC@NLO 2.2.3~\cite{Alwall:2014hca}} interfaced to \pythiaeight with
the NNPDF3.0LO~\cite{Ball:2014uwa} PDF set. Samples of $b\overline{b}h$ events are
generated by \amcnlo\_aMC@NLO 2.2.3 interfaced to \pythiaeight with the A14 tune and the NNPDF2.3LO
PDF set. The non-resonant diphoton processes with associated jets are generated using
\sherpa~\cite{Gleisberg:2008ta} with the CT10 PDF set. Matrix elements are
calculated with up to three partons at LO and merged with the \sherpa parton
shower~\cite{Schumann:2007mg} using the ME+PS@LO prescription~\cite{Hoeche:2009rj}. The CT10 PDF set is used in
conjunction with a dedicated parton-shower tuning developed by the \sherpa
authors. The $W\gamma$, $Z\gamma$, $W\gamma\gamma$, $Z\gamma\gamma$ samples are
generated using \sherpa with the CT10 PDF set.

The normalization of non-resonant backgrounds is obtained directly from data,
as described in Section~\ref{sec:backgroundest}.
The cross sections~\cite{deFlorian:2016spz} of the SM Higgs-boson processes are
calculated at next-to-leading order (NLO) in electroweak theory and
next-to-next-to-leading order (NNLO) in QCD for VBF,
$Zh$ and $Wh$ samples and next-to-next-to-next-to-leading order
plus next-to-next-to-leading logarithm (N$^{3}$LO+NNLL) in QCD for the ggF
sample. The SM Higgs-boson mass is set to 125.09~\GeV~\cite{Aad:2015zhl} and its branching fraction
decaying into two photons is 0.227\%~\cite{deFlorian:2016spz}.
Samples for the $Z'_{B}$ and $Z'$-2HDM models are normalized using the LO
cross sections predicted by \amcnlo\_aMC@NLO 2.2.3. Samples for the heavy-scalar
model are normalized using the cross section of a SM Higgs boson produced in ggF at
the same mass at NNLO+NNLL in QCD.

Different pileup conditions from same and neighboring bunch crossings as a function of the instantaneous luminosity are
simulated by overlaying minimum-bias events generated with \pythiaeight and
\evtgen~\cite{Lange:2001uf} with the MSTW2008LO PDF set and
the A2~\cite{ATLAS:2012uec} tune on the events of all hard processes.
Differences between the simulated and observed distributions of the number of
interactions per bunch crossing are removed by applying pileup weights to simulated events.
Detector effects are simulated using a full simulation~\cite{Aad:2010ah}
performed using \geantfour~\cite{Agostinelli:2002hh} for signals, SM Higgs
processes, and $W\gamma$, $Z\gamma$, $W\gamma\gamma$, and $Z\gamma\gamma$
backgrounds. The diphoton continuum background and some of
the signal samples are simulated using a fast simulation based on \atlfasttwo~\cite{Aad:2010ah}.

 \section{Event reconstruction}
\label{sec:reconstruction}

In each event, the observed final state is reconstructed from photons, leptons, jets, and \etmiss
that are built combining the related measurements provided by the various
subdetectors of the experiment.

Photons are reconstructed from clusters of energy deposits in the electromagnetic calorimeter measured in projective towers.
Clusters without matching tracks are classified as unconverted photon candidates.
A photon is considered as a converted photon candidate if it is matched to a pair of tracks that pass a TRT-hits
requirement~\cite{photonrecorunI} and that form a vertex in the ID which is consistent with originating from a massless particle,
or if it is matched to a single track passing a TRT-hits requirement and having a first hit after the innermost layer of the pixel detector.
The photon energy is calibrated using a multivariate regression algorithm
trained with fully reconstructed MC samples and then corrections based on in situ techniques on data, as explained in Ref.~\cite{Aad:2014nim}.
The overall energy scale in data, as well as the difference in the constant term of the energy
resolution between data and simulation, are estimated with a sample of $Z$ boson decays to electrons recorded in 2015 and 2016.
The photon direction is estimated either using electromagnetic (EM) calorimeter longitudinal
segmentation (if unconverted) or conversion vertex position (if converted), together with constraints from
the $pp$ collision point.

A ``tight'' photon identification requirement~\cite{photonrecorunI} is applied to reduce
the misidentification of hadronic jets containing a high-$\pt$ neutral hadron (e.g.~$\pi^0$) decaying to two photons.
The photon identification is based on the lateral profile of the energy
deposits in the first and second layers of the electromagnetic calorimeter, and
on the shower leakage fraction in the hadronic calorimeter.
The selection requirements are tuned separately for unconverted and converted photon candidates.
The identification efficiency of unconverted (converted) photons range from 85 to 95\% (90 to 98\%)
between 25 and 200\,\GeV~\cite{ATL-PHYS-PUB-2016-014}. Corrections are
applied to the electromagnetic shower-shape variables for simulated photons, to account for small differences observed between data and
simulation. The diphoton mass resolution at $\mgg = 125$~\GeV\ is in the range 1.32--1.86~\GeV~\cite{Aad:2014eha}.

To further reject hadronic backgrounds, requirements on two photon isolation variables are applied.
The first variable, $\et^\text{iso}$, is the sum of the transverse energies
deposited in topological clusters~\cite{Aad:2016upy} of cells in the calorimeter within a cone of 
size ${\Delta R \equiv \sqrt{(\Delta\eta)^2 + (\Delta\phi)^2}} = 0.2$ around each photon.
The photon cluster energy and an estimate of the energy deposited by the photon
outside its associated cluster are subtracted from that sum. 
To reduce the effects from the underlying event and pileup, the median ambient energy computed 
from low-$\pT$ jets in the event~\cite{Cacciari:2008gn,Aad:2015ina} is subtracted from $\et^\text{iso}$. 
Good candidates are required to have an $\et^\text{iso}$ less than 6.5\% of the photon transverse energy.
The second variable is a track-based isolation, defined as the scalar sum of the transverse momenta
of all tracks with $\pT>1$~\GeV\ and consistent with originating from the
primary vertex (PV) within a cone of size ${\Delta R = 0.2}$ around each photon. 
In the case of converted photon candidates, the tracks associated to the conversion are removed.
Good candidates are required to have a track isolation less than 5\% of the photon transverse energy.
The isolation efficiency for photons, which is mostly independent of their kinematic variables, is about
90\% for the ggF SM Higgs boson process.

Selected events contain at least one PV formed from two or
more tracks, each with $\pt > 400$~MeV. In each event, the diphoton PV is chosen 
from the PV candidates using a neural network. The input variables to this neural network are the combined $z$-position of
the intersections of the extrapolated photon trajectories
with the beam axis; the sum of the squared transverse momenta $\sum \pT^2$ and
the scalar sum of the transverse momenta $\sum \pT$ of the tracks associated
with each reconstructed vertex;
and the difference in azimuthal angle $\Delta \phi$ between the direction
defined by the vector sum of the momenta of tracks from each vertex and that
of the diphoton system. Dedicated studies of $Z\rightarrow e^{+}e^{-}$ are performed in order to validate
the diphoton vertex identification efficiency (correct identification of the hard process vertex by the neural network)
between data and simulation.
The method is similar to the one used in
Ref.~\cite{Aad:2014aba}. Studies show good agreement in diphoton vertex identification efficiency between data and simulation.
The efficiency to locate the diphoton PV within $0.3$~mm of the production
vertex is 81\% for SM Higgs boson production via ggF, 82\% for a heavy-scalar signal with $\mBH = 275$ GeV and scalar DM $m_\chi = 60$ GeV,
67\% for a $Z'_{B}$ signal with $m_{Z^{\prime}_{B}}=200~\GeV$ and Dirac fermion DM $m_{\chi}=1~\GeV$, 
and 69\% for a $Z'$-2HDM signal with $m_{Z^{\prime}}=1000~\GeV$, $m_{A^0} = 200~\GeV$ and Dirac fermion DM $m_{\chi}=100~\GeV$.

Electrons are reconstructed from energy deposits measured in the EM calorimeter
which are matched to ID tracks~\cite{Aaboud:2016vfy}. They are required to satisfy $|\eta| < 2.47$, excluding the EM calorimeter
transition region $1.37 < |\eta| < 1.52$, and to have $\pT > 10$ \GeV. The electrons are identified using
a likelihood-based algorithm that uses the track and shower-shape variables as input. The ``medium''
criteria are applied, providing an identification efficiency varying from 85 to 95\% as a function of \ET~\cite{ATL-PHYS-PUB-2015-041}.
Loose calorimeter and track isolation requirements are applied to electrons. 
Pileup and the underlying event contributions in the calorimeter isolation 
are estimated using the ambient energy from low-$\pT$ jets and are corrected on an event-by-event basis.
In the inclusive diphoton sample the combined efficiency of the isolation
requirements is 98\%~\cite{Aad:2014lwa}.

Muons are reconstructed from high-quality track segments in the muon
spectrometer. In the region $|\eta| < 2.5$, they must be matched to ID tracks. They are required to have $\pT > 10$~\GeV\ and $|\eta| < 2.7$~\cite{Aad:2014rra}.
The muon ``medium'' criteria are applied and the identification efficiency is 96\%~\cite{Aad:2016jkr}. The muon
candidates must also satisfy calorimeter and track isolation criteria. 
The ambient energy from low-$\pT$ jets is used to correct the contributions from pileup and the underlying event.
The combined isolation efficiency varies from 95\% to 99\% as a
function of \pT from 25~\GeV\ to 60~\GeV~\cite{Aad:2016jkr}.

The significance of the track's transverse impact parameter with respect to the diphoton primary vertex,
$|d_0|/\sigma_{d_0}$, is required to be less than 5.0 for electrons and 3.0 for muons. The longitudinal
impact parameter $z_0$ must satisfy $|z_0|\sin\theta <$ 0.5 mm for electrons and muons.

Jets are reconstructed from three-dimensional topological clusters using the anti-\kT algorithm~\cite{Cacciari:2008gp}
with a radius parameter of $R = 0.4$. The jets are required to have $\pT > 20$~\GeV\ and $\abseta < 4.5$ for
the \etmiss calculation and $\pT > 25$~\GeV\ and $\abseta < 4.4 $ for the event selection. The jets with $\abseta<2.4$
and $\pT < 60$~\GeV\ must pass the jet vertex tagger
selection~\cite{ATL-PHYS-PUB-2014-001}, in which a jet is identified as originating from the diphoton primary vertex
by examining the likelihood value calculated from the track information. In addition, quality criteria
are applied to the jets, and events with jets consistent with noise in the calorimeter or non-collision backgrounds are
vetoed~\cite{Aaboud:2017jcu}.

The missing transverse momentum is calculated as the magnitude of the negative vectorial sum of the transverse momenta
of calibrated photons, electrons, muons, and jets associated with the diphoton primary vertex. The transverse
momenta of all tracks that originate from the diphoton primary vertex
but are not already used in the \etmiss calculation are summed and taken into account in the \etmiss calculation.
This term is defined as
the track-based soft term~\cite{Aad:2012re, ATL-PHYS-PUB-2015-027}. Clusters and tracks not associated
with the diphoton primary vertex are not included in the \etmiss calculation, significantly suppressing the
pileup effect and thus improving the \etmiss resolution.

\section{Event selection}
\label{sec:evtselection}

Events are required to have at least two photon candidates with $\pT > 25$~\GeV\
and within a fiducial region of the EM calorimeter defined by $|\eta| < 2.37$, excluding the region of $1.37<|\eta|<1.52$.
Photon candidates in this fiducial region are ordered
according to their $E_\mathrm{T}$ and only the first two are considered: the leading and subleading photon candidates
must have
$E_\mathrm{T}^{\gamma}/m_{\gamma\gamma} > 0.35$ and 0.25, respectively, where $m_{\gamma\gamma}$ is the invariant mass of the two
selected photons.
Events are required to have $105 < m_{\gamma\gamma} < 160~\GeV$\, where the diphoton mass is calculated
assuming that the photons originate from the diphoton primary vertex.

In the $Z'_B$ and $Z'$-2HDM models of DM production, the Higgs boson recoils against the DM pair, resulting in large \etmiss\ in the event and
large \pT of the diphoton candidate, denoted as \pTgg. By contrast, in the heavy-scalar model, the spectra of \etmiss\ and \pTgg are typically shifted to smaller values.
Consequently, the classification of the selected events into categories depending
on \etmiss, \pTgg and other kinematic quantities leads to an increase in the
overall sensitivity of the analysis to these different signal models.

The background events that survive the high-\etmiss requirement but are not expected to
present any genuine \etmiss mostly have one or several high-\pT pileup jets. The
misidentification of these jets therefore leads to large \etmiss.
These pileup jets usually originate from a pileup vertex with larger $\Sigma\pt^{2}$ than the diphoton primary vertex,
where $\pt$ is the track transverse momentum associated with a single vertex.
Requiring the diphoton primary vertex to be the vertex with the largest
$\Sigma\pt^{2}$ in each event
helps to suppress pileup effects and reject a large fraction of the fake \etmiss events.
   Models for which ggF is the main production mode typically pass this selection,
 since for example the heavy scalar produced in ggF is often accompanied by
 radiated jets, leading to a large $\Sigma\pt^{2}$ of the vertex.
 An additional variable, \pThard, defined as the magnitude of the vectorial sum
of the transverse momenta photons and jets in the event, provides further discrimination against events with
fake \etmiss in the low-\etmiss region.
To further reject the background events from SM $V\gamma$ and $V\gamma\gamma$
production (where $V$ stands for $W$ or $Z$), which contribute significantly in the high-\etmiss region, a lepton (electron or muon) veto is applied.

The selected events are thus divided into five categories based on:
\begin{itemize}
	\item the \etmiss significance, $\setmiss = \etmiss / \sqrt{\sum E_\mathrm{T}}$,
	where the total transverse energy $\sum E_\mathrm{T}$ is calculated from the scalar sum of the transverse momenta of the calibrated photons,
	electrons, muons and jets used in the \etmiss calculation described in Section~\ref{sec:reconstruction}, as well as the tracks not associated with these but the PV;
	\item the diphoton transverse momentum, \pTgg;
	\item \pThard;
	\item the number of leptons in the event;
	\item the diphoton vertex is the highest $\Sigma\pT^{2}$ vertex.
\end{itemize}

The resulting categorization scheme is shown in Table~\ref{tab:cat_final}. The
categories are defined sequentially in the rows and each category excludes
events in the previous row.
The $Z^\prime_B$ and $Z^\prime$-2HDM signal samples are used to optimize the definition of the Mono-Higgs category,
which provides most of the sensitivity to those two models. The other four
categories, which provide extra sensitivity to heavy scalar boson events with softer \etmiss, are optimized using simulated heavy scalar boson samples to
cover the different kinematic regimes of the heavy-scalar model.

\begin{table*}[!htbp]
  \caption{Optimized criteria used in the categorization. The categories are
  defined sequentially in the rows and each category excludes
  events in the previous row.
          }
  \begin{center}
    \begin{tabular}{ll}
      \toprule
      \toprule
      Category & Requirements  \\
      \midrule
      Mono-Higgs  & $ \setmiss > 7\ \sqrt{\textrm{GeV}}$, $\pTgg > 90$ \GeV, lepton veto \\
      High-\etmiss  & $ \setmiss >  5.5\ \sqrt{\textrm{GeV}}$, $\mathrm{PV}^\text{highest} = \mathrm{PV}^{\gamma\gamma}$ \\
      Intermediate-\etmiss  & $  \setmiss > 4\ \sqrt{\textrm{GeV}}$, $\pThard > 40$~\GeV , $\mathrm{PV}^\text{highest} = \mathrm{PV}^{\gamma\gamma}$  \\
      Different-Vertex   & $  \setmiss > 4\ \sqrt{\textrm{GeV}}$, $\pThard > 40$~\GeV , $\mathrm{PV}^\text{highest} \neq \mathrm{PV}^{\gamma\gamma}$  \\
      Rest &$ \pTgg > 15 $~GeV \\
       \bottomrule
       \bottomrule
    \end{tabular}
    \label{tab:cat_final}
  \end{center}
\end{table*}

\begin{figure}[!htpb]
\centering
   \begin{tabular}{cccc}
	\subfigure[]{\includegraphics[width=0.49\textwidth]{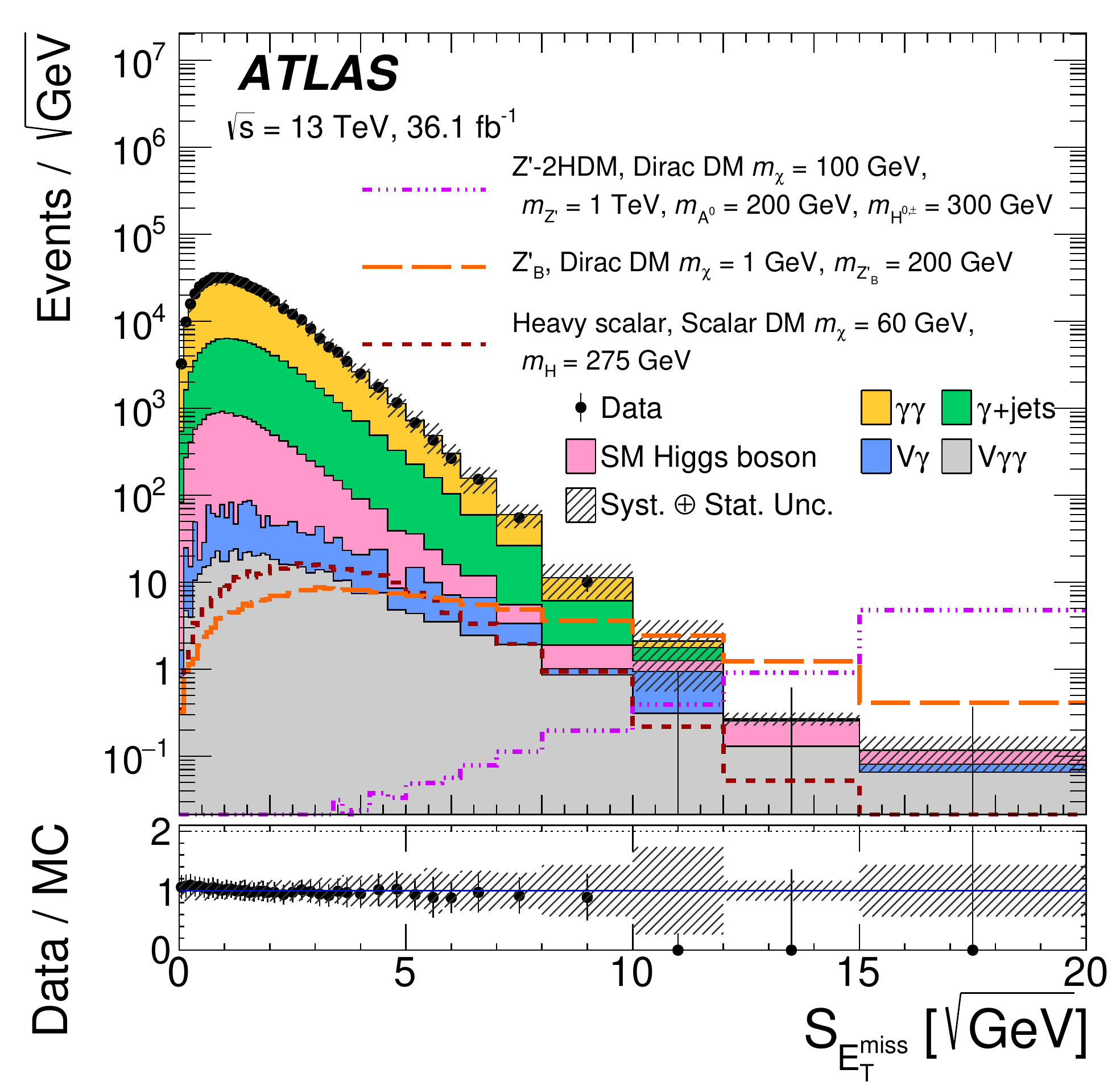}}
	\subfigure[]{\includegraphics[width=0.49\textwidth]{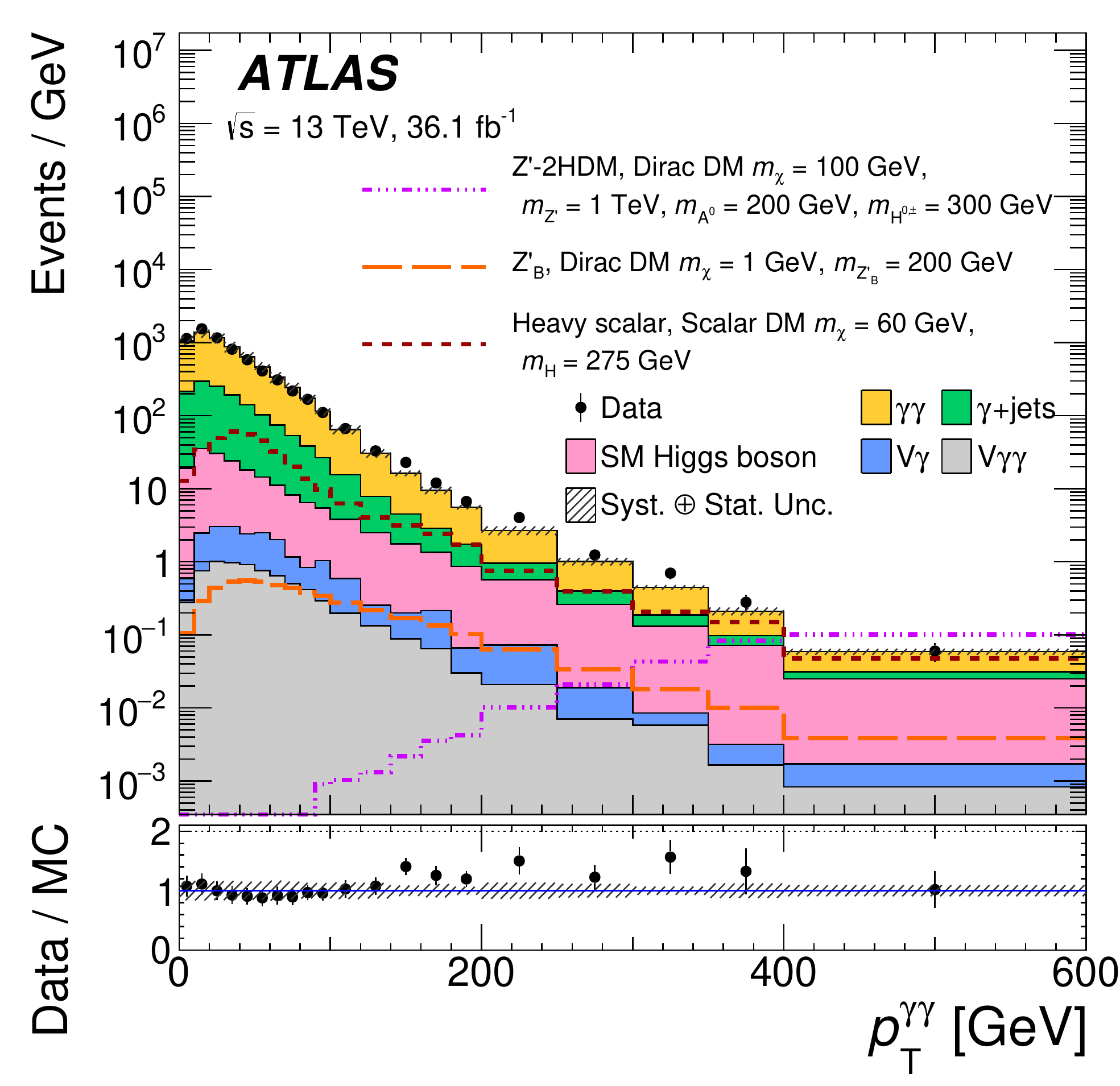}} \\
	\subfigure[]{\includegraphics[width=0.49\textwidth]{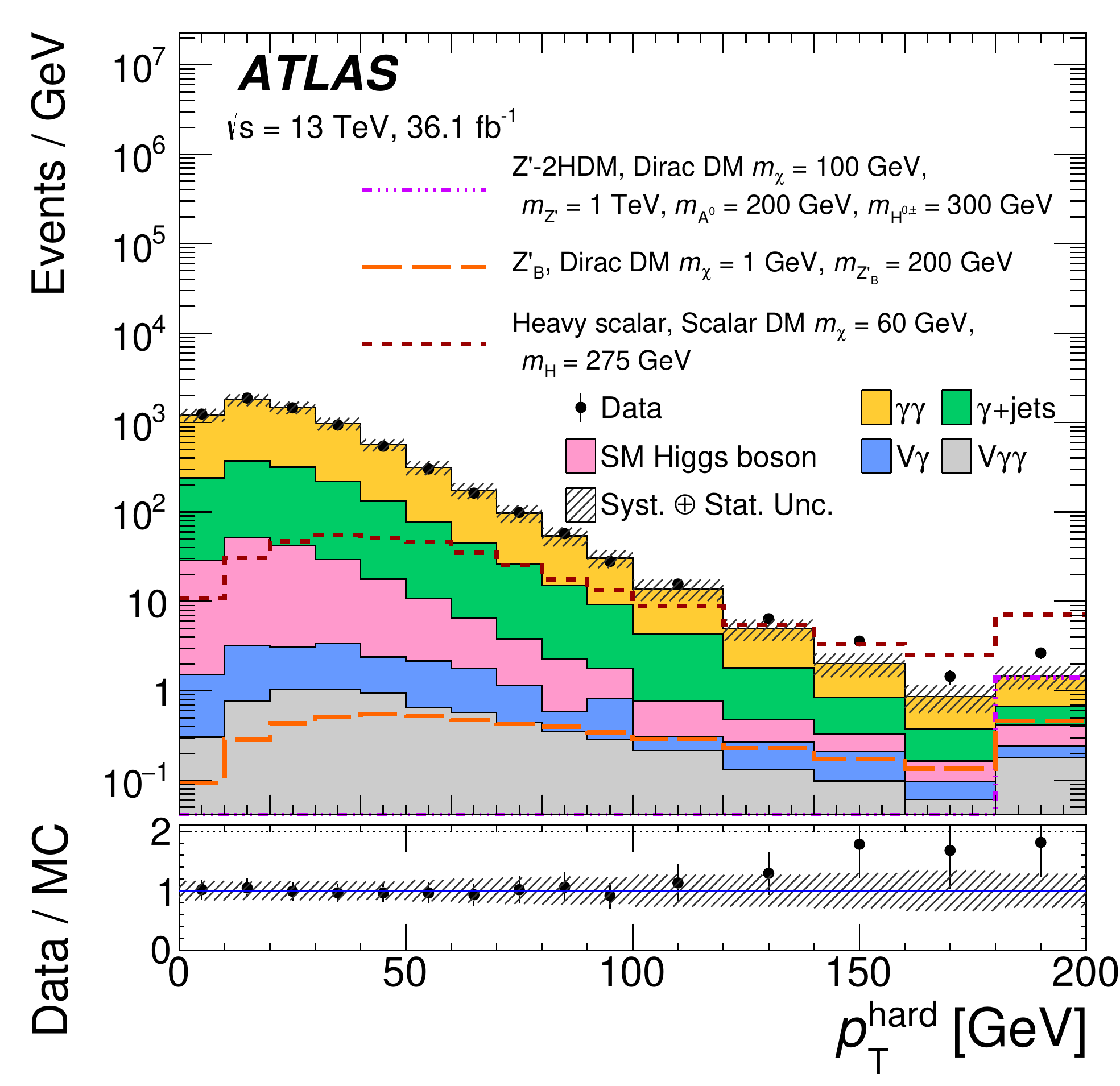}}
    \end{tabular}
    \caption{
    	The distribution of (a) \setmiss, (b) \pTgg, and (c) \pThard after the selection of diphoton candidates
	in \mbox{$120 < m_{\gamma\gamma} < 130~\GeV$}. Expected distributions are shown
	for a $Z'_{B}$ signal with $m_{Z^{\prime}_{B}}=200~\GeV$ and Dirac
        fermion DM $m_{\chi}=1~\GeV$; a $Z'$-2HDM signal with
        $m_{Z^{\prime}}=1000~\GeV$, $m_{A^0} = 200~\GeV$ and Dirac fermion DM
        $m_{\chi}=100~\GeV$; and a heavy-scalar model with $\mBH = 275$ GeV and scalar DM $m_\chi = 60$ GeV.
        	These overlaid signal points are representative of the model kinematics.
        Only the quadratic sum of the MC statistical and experimental systematic uncertainties in the total background is shown as the hatched bands,
        while the theoretical uncertainties in the background normalization are not included.
        			Overflow events are included in the rightmost bin.
	The asymmetric error bars on data points come from Poissonian confidence intervals
	at 68\% confidence level.
	}
   \label{fig:controls}
\end{figure}

Figure~\ref{fig:controls} shows the distributions of \setmiss, \pThard, and \pTgg after the selection of diphoton candidates
in $120 < m_{\gamma\gamma} < 130~\GeV$. 
These distributions illustrate three kinds of signals in different \setmiss regimes, and the contributions from the different background processes.
Expected distributions are shown for a $Z'_{B}$ signal
with $m_{Z^{\prime}_{B}}=200~\GeV$ and Dirac fermion DM $m_{\chi}=1~\GeV$, a $Z'$-2HDM signal with $m_{Z^{\prime}}=1000~\GeV$,
$m_{A^0} = 200~\GeV$ and Dirac fermion DM $m_{\chi}=100~\GeV$, and a heavy-scalar model with $\mBH = 275$ GeV and scalar DM $m_\chi = 60$ GeV. These overlaid signal points are representative of the model kinematics.

For the distributions shown in Fig.~\ref{fig:controls}, the simulation is used to determine the shapes and
normalizations of the $V\gamma$ and $V\gamma\gamma$ contributions, as well as
the shape of the $\gamma\gamma$ contribution, respectively.
The normalizations of the $\gamma\gamma$ and $\gamma$+jet contributions are
fixed to 79 and 19\% of the data yield, where these fractions are estimated from a
two-dimensional sideband technique by counting the number of events in which one
or both photons pass or fail the identification or isolation
requirements~\cite{Aad:2012tba}.
The shape of the $\gamma$+jet contribution is determined from a data control region
where one photon fails to satisfy the photon identification criteria, after subtracting the
contamination expected from $\gamma\gamma$, $V\gamma$ and $V\gamma\gamma$.

The slight discrepancies observed in the distributions of \setmiss, $\pTgg$, and
$\pThard$ in Fig.~\ref{fig:controls} do not affect the results. Discrepancies
are found mainly in non-resonant backgrounds, which are estimated directly from data, as explained in Section~\ref{sec:backgroundest}.

 \section{Signal and background parameterization}
\label{sec:backgroundest}

The signal and backgrounds are extracted by fitting analytic functions to
the diphoton invariant mass distribution in each category.
For the signal and the SM Higgs-boson background, the expected normalizations
are obtained from their theoretical cross sections multiplied by the product of
the acceptance times efficiency from the simulation.
The shapes are modeled with a double-sided
Crystal Ball function (as defined in Ref.~\cite{Aad:2014eha}). The shape
parameters are determined by fitting the diphoton mass distribution in simulation for each category.

Both the normalization and the shape of the non-resonant background are obtained
by fitting the diphoton invariant mass distribution in data for each category.
A variety of analytic functions are considered for the non-resonant background parameterization:
exponential functions of different-order polynomials, Bernstein polynomials of
different order, and an adapted dijet function~\cite{Aaltonen:2008dn}.
The potential bias associated with the choice of a specific analytic function
to model the continuum background
(referred to as the non-resonant background modeling uncertainty, $\Delta N^\mathrm{non-res}_\mathrm{bkg}$ ) is estimated for each category as the
signal event yield extracted from a signal-plus-background maximum-likelihood fit to 
a background-only diphoton invariant mass distribution with small statistical fluctuations~\cite{Aad:2014eha}.
The background-only distribution is obtained by mixing simulated $\gamma\gamma$, $\gamma$+jet, $V\gamma$, and $V\gamma\gamma$
processes. The samples of $V\gamma$ and $V\gamma\gamma$ are
weighted according to their theoretical cross section while $\gamma\gamma$ and $\gamma$+jet samples
are normalized to the number of candidates in data in the mass window $105 < \mgg < 160~\GeV$ scaled by the fraction of each
sample (79\% for $\gamma\gamma$ and 19\% for $\gamma$+jet).
For a given functional form, several fits are tested by varying the position
of the signal peak between $115$ and $135$~\GeV.
The largest number of signal events obtained in these fits to the background-only templates is taken as $\Delta N^\mathrm{non-res}_\mathrm{bkg}$.
Among the different analytic functions that were tested, the parameterization with the smallest $\Delta N^\mathrm{non-res}_\mathrm{bkg}$, or the minimum
number of free parameters when the same $\Delta N^\mathrm{non-res}_\mathrm{bkg}$ values are obtained, 
is selected as the nominal background parameterization to describe the non-resonant background shape.
In addition, a $\chi^{2}$ test is performed to ensure that the fit is compatible
with the data in each category.

The sidebands ($105 < \mgg < 120~\GeV$ and $130 < \mgg < 160~\GeV$) of data and the simulated events are compared with each other. 
Due to the low statistics in the Mono-Higgs category, the potential discrepancies between data and the simulated events 
are obscured by the statistical uncertainties.
A visible discrepancy is only observed in the High-\etmiss, Intermediate-\etmiss, Different-Vertex, and Rest categories with large data statistics
and in these cases a data-driven reweighting is applied to the simulated events to correct the shape.
In order to check whether there is a substantial improvement in $\chi^{2}$ between two nested functions, 
several $F$-tests~\cite{Aaboud:2016tru} are performed in the data sidebands 
of the Mono-Higgs category where the simple exponential function is chosen. 
The $F$-tests are performed by comparing the simple exponential function to other higher-order functions.
A test statistic $F$ is calculated by using the resulting values of $\chi^{2}$, 
and its probability is compared with that expected from a Fisher distribution with the corresponding number of degrees of freedom.
The hypothesis that a higher-order function is not needed is rejected if the $F$-test probability is less than 5\%. 
The tests show that the higher-order functions don't provide a significantly better fit and the lower-order function is sufficient.
The selected analytic function along with its non-resonant background modeling
uncertainty in each category, which is taken as an estimate of the systematic
uncertainty due to the choice of parameterization, is shown in Table~\ref{tab:decision}.

\begin{table*}[!htbp]
  \caption{The analytic functions used to model the non-resonant \mgg
  distribution, the uncertainty in the signal due to the non-resonant background
  modeling ($\Delta N^\mathrm{non-res}_\mathrm{bkg}$) and the relative
  uncertainties in the expected non-resonant background and signal per category.
   The relative uncertainty in the signal is evaluated in the Mono-Higgs category
 by using the $Z'_B$ yields and in the other categories by using the heavy-scalar yields as $N_{\mathrm{signal}}$ from Table~\ref{tab:cut}. 
 The signal yields are calculated in the \mgg range between $120$ and $130$~\GeV.
     The variable $x$ is defined as $m_{\gamma\gamma}/\sqrt{s}$ while $a$ and $b$ are parameters of the background functions. For the Rest category, $C_3^j$
are binomial coefficients and $b_{j,3}$ are Bernstein coefficients. } 
\begin{tabular}{ll...}
      \toprule
      \toprule
      \multicolumn{1}{l}{ Category } &  \multicolumn{1}{l}{ Function}  &  \multicolumn{1}{c}{ $\Delta N^\mathrm{non-res}_\mathrm{bkg}$ } &
      \multicolumn{1}{c}{ $\Delta N^\mathrm{non-res}_\mathrm{bkg}$/$N^{\mathrm{non-res.}}_{\mathrm{bkg}}$ [\%]} &
		      \multicolumn{1}{c}{ $\Delta N^\mathrm{non-res}_\mathrm{bkg}$/$N_{\mathrm{signal}}$ [\%]} \\
      \midrule
      Mono-Higgs  &  $\exp(a\cdot x)$ & 1.2 & 9.8 & 6.0 \\
      High-\etmiss  &   $(1-x^{1/3})^b \cdot x^a$ & 2.7 & 4.0 & 11 \\
      Intermediate-\etmiss  & $\exp(a\cdot x+b\cdot x^{2})$  & 5.8 & 1.3 & 14\\
      Different-Vertex   &   $\exp(a\cdot x+b\cdot x^{2})$ & 8.4 & 0.5 & 26 \\
      Rest &  $\sum_{j=0}^3 C_3^j x^j(1-x)^{3-j} b_{j,3}$ & 61 & < 0.1 & 28 \\
       \bottomrule
       \bottomrule
    \end{tabular}
    \label{tab:decision}
\end{table*}

 \section{Systematic uncertainties}
\label{sec:effsys}

Uncertainties from experimental and theoretical sources affect the signal
efficiency and the SM Higgs-boson background yield estimated from the simulated
MC samples.
The non-resonant background is obtained directly from the fit to the data.
The only systematic uncertainty in the non-resonant background is the potential
bias $\Delta N^\mathrm{non-res}_\mathrm{bkg}$.
A summary of the experimental and theoretical uncertainties with respect to the yield of
the background from SM Higgs-boson processes, non-resonant background, and
signal production is shown in Table~\ref{tab:unc}.

\begin{table*}[!htp]
  \caption{Breakdown of the dominant systematic uncertainties in the range of 105 $<\mgg <$ 160 \GeV.
	  The uncertainties in the yield of signals, the background
	  from SM Higgs-boson processes, and non-resonant background are shown.
	  All production modes of the SM Higgs boson are considered together. 	  Values for the impact on all categories are shown, unless one of the
	  systematic uncertainties is not applicable to the sample, in which case
	  the value is substituted by a ``-''. If a given source has a different
	  impact on the various categories, the given
	  range corresponds to the smallest and largest impacts
	  among categories or among the different signal models used
	  in the analysis. In addition, the potential bias coming from non-resonant
	  background mismodeling is shown relative to the background.
          }
\begin{center}
\begin{tabular}{l..c}
      \toprule
      \toprule
      \multirow{3}{*}{Source}           & \multicolumn{1}{c}{\multirow{3}{*}{Signals [\%]}}  & \multicolumn{2}{c}{Backgrounds [\%]} \\
      \cline{3-4}
       &  & \multicolumn{1}{c}{SM Higgs boson} & \begin{tabular}{c} Non-resonant \\  background \end{tabular} \\
      \midrule
      \multicolumn{3}{l}{Experimental} \\
      \midrule
      Luminosity & \multicolumn{2}{c}{3.2}  & - \\        Trigger efficiency & \multicolumn{2}{c}{0.4} & - \\
      Vertex selection  & \multicolumn{2}{c}{< 0.1} & - \\
      Photon energy scale & 0.1-2.0 & 0.1-1.4  & - \\
      Photon energy resolution & 0.1-0.2 & 0.1-1.1  & -\\
      Photon identification efficiency & 2.9-4.3 & 1.9-3.8   & - \\
      Photon isolation efficiency & 1.2 & 0.8-1.6  & - \\
      \etmiss\ reconstruction (diphoton vertex) & < 0.1  &  0.5-1.9 & - \\
      \etmiss\ reconstruction (jets, soft term) & 1.0-1.4 &  0.8-23   & - \\
      Diphoton vertex with largest $\Sigma\pt^{2}$ & < 0.1 -1.9 & < 0.1-6.0 & - \\
      Pileup reweighting & 0.2-5.6  & 0.7-11 & -\\
      Non-resonant background modeling & - & - &   $0.1-9.8$  \\
      \midrule
      \multicolumn{3}{l}{Theoretical} \\
      \midrule
      Factorization and renormalization scale & 0.6-11 & 2.5-6.0  & - \\
      PDF+$\alpha_\mathrm{S}$ & \multicolumn{1}{c}{$11-25$} & 1.2-2.9   & - \\
      Multiple parton-parton interactions & <1  & 0.4-5.8 & - \\
      $\cal{B}$($H\rightarrow \gamma\gamma$) & \multicolumn{2}{c}{1.73} & - \\
       \bottomrule
       \bottomrule
    \end{tabular}

    \label{tab:unc}
 \end{center}
\end{table*}

The uncertainty in the combined 2015 and 2016 integrated luminosity is 3.2\%.
It is derived, following a methodology similar to that detailed in
Ref.~\cite{Lumi2}, from a calibration of the luminosity scale
using $x$--$y$ beam-separation scans performed in August 2015 and May 2016.

The efficiency of the diphoton trigger used to select events is evaluated in MC
simulation using
a trigger matching technique and in data using a bootstrap method~\cite{Aaboud:2016leb}.
In the diphoton invariant mass window of $105 < m_{\gamma\gamma} < 160~\GeV$, the trigger efficiency
uncertainty is found to be 0.4\%.

The uncertainty in the vertex selection efficiency is assessed by comparing the efficiency of finding
photon-pointing vertices in $Z \rightarrow e^+e^-$ events in data and MC
simulation~\cite{Aaboud:2016tru}, for which each electron track is removed and its cluster
is recalibrated as a photon cluster.
The efficiency of this selection in data is found to be in agreement with the simulation within 0.01\%.

The systematic uncertainties due to the photon energy scale and
resolution are adapted from Run-1 results~\cite{Aad:2014nim}, with minor
updates in case of data-driven corrections
using the Run-2 data. The uncertainty in the energy scale is less than $1\%$ in
the \pT range of the photons used in this analysis; the uncertainty in the energy resolution is smaller than $2\%$.

Uncertainties in photon identification and isolation efficiencies are estimated, and their impact on the number of events in each category is quantified.
The uncertainty in the photon identification efficiency~\cite{ATL-PHYS-PUB-2016-014} is calculated
from the difference between applying and not applying the corrections to the
electromagnetic-shower-shape variables in simulation.
The resulting uncertainty in the photon identification efficiency
is lower than $3.8\%$ for SM Higgs background in all categories,
$2.9\%$ for simplified model samples, and $4.3\%$ for the heavy-scalar model.
The uncertainty in the photon calorimeter isolation efficiency is calculated from efficiency differences
between applying and not applying corrections derived from inclusive photon events to the isolation variables in simulation.
The measurements of the efficiency correction factors using inclusive photon
events are used to derive the efficiency uncertainty in the photon track isolation uncertainty.
The photon isolation efficiency uncertainty is found to be
smaller than $1.6\%$ for the SM Higgs background and $1.2\%$ for all signal samples.

Migration of events among categories occurs owing to
changes in the energy of identified particles and jets, mostly due to the misreconstruction of jets and \etmiss.
The uncertainties in jet energy scale, resolution, and jet vertex
tagger are propagated to the \etmiss calculation. In addition, the
uncertainties in the scale and resolution of the \etmiss soft term are estimated in 2016 data using the method
described in Ref.~\cite{ATL-PHYS-PUB-2015-023}.
The overall jet and \etmiss uncertainties in the SM Higgs-boson processes are 6, 8, 23, 22, and 1\% for the Mono-Higgs, High-\etmiss,
Intermediate-\etmiss, Different-Vertex, and Rest categories, respectively.
For signal processes, the overall jet and \etmiss uncertainties range from 1.0 to 1.4\%.
The choice of the diphoton vertex also affects the \etmiss reconstruction. It introduces
an additional uncertainty derived from the data-to-MC comparison in $Z
\rightarrow e^+e^-$ events. This systematic uncertainty affects the processes
with no genuine \etmiss and is estimated in each category. For the SM
Higgs-boson production, it is found to be 0.5\% in the Mono-Higgs category and
up to 1.9\% in the other categories. It is less than 0.1\% for signal processes.
Requiring the diphoton primary vertex to be the vertex with the largest $\Sigma\pt^{2}$ in
the event introduces an uncertainty of about 6\% in the SM Higgs-boson production
in the High-\etmiss and the Intermediate-\etmiss categories and
1.8\% in the heavy-scalar signals in those categories.
This systematic uncertainty differs between the signal
and background because there is a correlation between
the \etmiss and the pileup vertex $\Sigma\pt^{2}$ for different processes.
For other signal samples, these uncertainties are at most a few percent in any category. The pileup reweighting uncertainty is taken into account by propagating it
through the event selection, and results in a 0.2--5.6\% uncertainty in the event yield of the signal and
SM Higgs-boson samples.

The non-resonant background contribution
is not affected by the same systematic uncertainties that characterize the MC simulation,
since it is extracted from the data.
The potential bias is added as a systematic uncertainty to account for the possible impact on the fitted signal yields
of non-resonant background mismodeling.
It is the leading systematic uncertainty of the analysis but is only one third
as large as the statistical uncertainty.
 The ratio of the potential bias in the signal yield to the expected non-resonant background 
($\Delta N^{\mathrm{non-res.}}_{\mathrm{bkg}}/N^{\mathrm{non-res.}}_{\mathrm{bkg}}$) in the range $120 <\mgg < 130~\GeV$ 
is 9.8\% in the Mono-Higgs category, 4.0\% in the High-\etmiss category, 1.3\% in the Intermediate-\etmiss category,
0.5\% in the Different-Vertex category, and 0.1\% in the Rest category.

The predicted cross sections of the SM Higgs-boson and signal processes are affected by uncertainties due to missing
higher-order terms in perturbative QCD. These uncertainties are estimated by varying the factorization and renormalization scales up and
down from their nominal values by a factor of two, recalculating the cross section in each case, and taking the largest deviation from
the nominal cross section as the uncertainty. The uncertainty related to the renormalization and factorization scales
is 0.6--11\% for signal and 2.5--6.0\% for the SM Higgs-boson processes~\cite{deFlorian:2016spz}.
 For the three signal processes, the effect of PDF+$\alpha_\mathrm{S}$
uncertainties including acceptance and normalization is 11--25\%. It is estimated using the recommendations
of PDF4LHC~\cite{deFlorian:2016spz}. Both intra-PDF and inter-PDF uncertainties are extracted.
Intra-PDF uncertainties are obtained by varying the parameters of the NNPDF3.0LO PDF set, while inter-PDF uncertainties
are estimated using alternative PDF sets (CT14~\cite{Dulat:2015mca} at LO and MMHT2014~\cite{Harland-Lang:2014zoa} at LO).
The final inter-PDF uncertainty is the maximum deviation among all the variations from the central
value obtained using the NNPDF3.0LO PDF set.
In the case of the SM Higgs-boson processes, the effect of $\alpha_\mathrm{S}$
and the choice of PDFs range from 2 to 6\%, which are taken from Ref.~\cite{deFlorian:2016spz}.
The uncertainty in the branching fraction ($\cal{B}$) of $h\rightarrow \gamma\gamma$ is 1.73\%~\cite{deFlorian:2016spz}.
The uncertainty in the multiple parton--parton interactions is estimated by switching them on and off in \pythiaeight\
in the production of the ggF SM Higgs-boson sample. The resulting uncertainty in the number of events in this sample
conservatively reaches $0.4\%$ in the Rest category, $5.8\%$ in the
Different-Vertex category, $3.8\%$ Intermediate-\etmiss and Different-Vertex
category, $3.4\%$ in the High-\etmiss category, and $1.4\%$ in the Mono-Higgs category.

\section{Results}
\label{sec:result}

The results for the analysis are derived from a likelihood fit of the \mgg
distribution in the range $105 < \mgg < 160~\GeV$. The SM Higgs boson
mass is set to $125.09$~\GeV~\cite{Aad:2015zhl}.
The impact due to the SM Higgs-boson mass uncertainty is negligible.
The signal strength, the background shape parameters, and the nuisance parameters representing the
systematic uncertainties are set to be free parameters. The SM Higgs yields are taken from the SM predictions.
The systematic uncertainty of
each nuisance parameter is taken into account by multiplying the likelihood by a Gaussian penalty function centered on the nominal value of this parameter with
a width set to its uncertainty. The nominal value of each SM Higgs-boson background nuisance parameter (including its yield) is taken from the simulation normalized to the SM theoretical predictions.
Since adding all the other categories to the Mono-Higgs category only brings a
1\% sensitivity improvement for both the $Z'_B$ and $Z'$-2HDM signals, the results are
only obtained from this category for these two simplified models. In contrast,
results for the heavy-scalar model are obtained from a simultaneous fit of all the categories.

The event yields in the observed data, expected signal and backgrounds in the five categories within a
window of \mbox{$120 < \mgg < 130~\GeV$} are shown in
Table~\ref{tab:cut}. The signal samples shown correspond to
a $Z'_B$ signal with $m_{Z^{\prime}_B}=200~\GeV$ and $m_{\chi}=1~\GeV$,
a $Z'$-2HDM signal with $m_{Z^{\prime}}=1000~\GeV$, $m_{A^0} = 200~\GeV$, and $m_{\chi}=100~\GeV$,
and a heavy-scalar signal with $\mBH=275~\GeV$ and $m_{\chi} = 60~\GeV$. For each benchmark signal model, the selection
efficiency times acceptance denoted by ``$\cal{A} \times \epsilon$'' is also
shown. The yields for the non-resonant background are obtained from a fit to data while
SM Higgs-boson events are estimated from the simulation. The uncertainties
correspond to the quadrature sum of the statistical and systematic uncertainties.

\begin{table*}[!htbp]
\small
  \caption{Event yields in the range of 120 $<\mgg <$ 130 \GeV\  for
  data, signal models, the SM Higgs-boson background and non-resonant background
  in each analysis category, for an integrated luminosity of $36.1$~fb$^{-1}$.
The signal samples shown correspond to a heavy-scalar sample with $\mBH=275~\GeV$ and $m_{\chi} = 60~\GeV$,
a $Z'_B$ signal with $m_{Z^{\prime}_B}=200~\GeV$ and $m_{\chi}=1~\GeV$ and
a $Z'$-2HDM signal with $m_{Z^{\prime}}=1000~\GeV$, $m_{A^0} = 200~\GeV$, $m_{H^{0,\pm}} = 300~\GeV$
and $m_{\chi}=100~\GeV$. For each benchmark signal model, the selection
efficiency times acceptance denoted as ``$\cal{A} \times \epsilon$'' is also shown.
The yields for non-resonant background are obtained from a fit to data while SM
Higgs-boson events are estimated from the simulation. The uncertainties
correspond to the quadrature sum of the statistical and systematic uncertainties. }
  \begin{center}
    \begin{tabular}{l.cccc}
      \toprule \toprule
      Category & \multicolumn{1}{c}{Mono-Higgs} & High-\met & Intermediate-\met & Different-Vertex & Rest \\
\midrule
Data  & 9 & 72 & 464 & 1511 & 46804 \\
\midrule
        \multicolumn{6}{c}{Backgrounds} \\
\midrule
SM Higgs boson  & 2.43 \pm 0.22 & $4.2 \pm 0.6$ & $11.9 \pm 2.7$ & $44 \pm 10$ & $1360 \pm 110$\\
Non-resonant  & 9.9 \pm 1.9 & $62 \pm 5$ & $418 \pm 10$ & $1490 \pm 18$ & $45570 \pm 110$\\
Total background  & 12.3 \pm 1.9 & $67 \pm 5$ & $430 \pm 10$ & $1535 \pm 21$ & $46930 \pm 170$\\
\midrule
	\multicolumn{6}{c}{$Z'_B$ model, $m_{Z^{\prime}_B}=200~\GeV$, $m_{\chi}=1~\GeV$} \\
\midrule
Expected yields				& 20.0 \pm 4.5 & $-$ & $-$ & $-$ & $-$ \\
$\cal{A} \times \epsilon$ [$\%$]		& 17.4 \pm 0.2  & $-$ & $-$ & $-$ & $-$ \\
\midrule
 	\multicolumn{6}{c}{$Z'$-2HDM model, $m_{Z^{\prime}}=1000~\GeV$, $m_{A^0} = 200~\GeV$, $m_{H^{0,\pm}} = 300~\GeV$, and $m_{\chi}=100~\GeV$}\\
\midrule
Expected yields				& 28.0 \pm 5.3 & $-$ & $-$ & $-$ & $-$ \\
$\cal{A} \times \epsilon$ [$\%$]		& 70.7 \pm 0.2 & $-$ & $-$ & $-$ & $-$ \\
\midrule
        \multicolumn{6}{c}{Heavy-scalar model, $\mBH=275~\GeV$, $m_{\chi}=60~\GeV$}\\
\midrule
Expected yields  & 10.9 \pm 1.4 & $23.8 \pm 3.2$ &  $43 \pm 5$ & $33 \pm 5$ & $222 \pm 20$\\
$\cal{A} \times \epsilon$ [$\%$]  & 1.22 \pm 0.07 & $2.67 \pm 0.10$ & $4.82 \pm 0.14$ & $3.65 \pm 0.13$ & $24.9 \pm 0.4$\\
       \bottomrule \bottomrule
    \end{tabular}
    \label{tab:cut}
  \end{center}
\end{table*}

\subsection{Limits on the visible cross section}

The observed yields agree with the SM background predictions, as shown in Table~\ref{tab:cut},
and no significant excess of events is observed.
Upper limits are set on the visible cross section $\sigma_{\mathrm{vis}}^{\mathrm{BSM}} \equiv (\cal{A} \times \epsilon \times \sigma \times {\cal{B}})^\mathrm{BSM}$
for BSM physics processes producing missing transverse momentum and a SM
Higgs boson decaying into two photons.
The SM background prediction is excluded from this BSM visible cross section.
Table~\ref{tab:visiblexsection} shows the observed and expected 95\% confidence level (CL) upper
limits on $\sigma_{\mathrm{vis}}^{\mathrm{BSM}}$, which are calculated using a one-sided profile-likelihood ratio and the CL$_\mathrm{s}$
formalism~\cite{Read:2002hq,Junk:1999kv} with the asymptotic approximation in Ref.~\cite{Cowan:2010js}.
The same parameterizations for the BSM signal
and the total SM Higgs-boson background are used in each of the five different
categories. The fits are performed individually in each category.
The statistical uncertainty is dominant.
The systematic uncertainties worsen the limits by about 10\% (7\% from the non-resonant background modeling and 3\% from the other systematic uncertainties).
The $\pm1\sigma$ variations from the expected limits are also given.
For the Mono-Higgs category, visible cross sections
$\sigma_{\mathrm{vis}}^{\mathrm{BSM}} > 0.19$~fb are excluded.
The ranges of the acceptance times efficiency ($\cal{A} \times\epsilon$) for
all three different models considered in this paper are also shown.
For the $Z'_{B}$ model, signals with DM mass $m_\chi$
between $1$ and $1000$~\GeV\ and mediator mass $m_{Z'_{B}}$ between $1$ and $2000$~\GeV\ are taken into consideration.
The samples with the mediator mass $m_{Z'}=400$--$1400$~\GeV\ and pseudoscalar boson mass
$m_{A^0}=200$--$450$~\GeV\ are added for the $Z'$-2HDM model.
For the heavy-scalar model, the values are taken from the signals points
with \mBH = 260 to 350 GeV and $\mchi = 60$~\GeV.

\begin{table*}[!htbp]
  \setlength{\extrarowheight}{2pt}
  \caption{
  	  Observed and expected upper limits (at 95\% CL) on the visible cross section for
	  BSM physics processes producing missing transverse momentum and a SM
	  Higgs boson decaying into two photons. Limits are presented for the five
	  different categories. The $\pm1\sigma$ exclusion from the expected limits
	  are also given. For all the simulated signal points, the lowest and largest
	  values of the acceptance times efficiency ($\cal{A} \times\epsilon$) for all three
  	  models are presented as a range. For the $Z'_{B}$ model, signals with DM mass $m_\chi$
	  between $1$ and $1000$~\GeV\ and mediator mass $m_{Z'_{B}}$ between $1$ and $2000$~\GeV\ are taken into consideration.
	  The samples with the mediator mass $m_{Z'}=400$--$1400$~\GeV\ and pseudoscalar boson mass
	  $m_{A^0}=200$--$450$~\GeV\ are added for the $Z'$-2HDM model.
	  For the heavy-scalar model, the values are taken from the signals points
	  with \mBH = 260 to 350 GeV and $\mchi = 60$~\GeV.
  }
  \begin{center}
        \begin{tabular}{l..ccc}
       \hline
       \hline
       \multirow{2}{*}{Category} & \multicolumn{2}{c}{$\sigma_{\mathrm{vis}}^{\mathrm{BSM}}$ [fb]} & \multicolumn{3}{c}{$\cal{A} \times\epsilon$ [\%]} \\ [0.5ex]
       \cline{2-3} \cline {4-6}
       &  \multicolumn{1}{c}{Observed} 	& \multicolumn{1}{c}{Expected} & \multicolumn{1}{c}{$Z'$-2HDM} & \multicolumn{1}{c}{$Z'_{B}$} & \multicolumn{1}{c}{Heavy scalar}\\\hline
       \multicolumn{1}{l}{Mono-Higgs} 			& 0.19 	& 0.23^{+0.11}_{-0.07} & \multicolumn{1}{c}{$53-74$} & \multicolumn{1}{c}{$15-63$} & $1.0-4.0$\\ [0.6ex]
       \multicolumn{1}{l}{High-\etmiss} 		& 0.67 	& 0.52^{+0.23}_{-0.15} & $0.2-12$ & $1.3-7.1$ & $1.8-8.4$\\ [0.6ex]
       \multicolumn{1}{l}{Intermediate-\etmiss} 	& 1.6  	& 1.2^{+0.5}_{-0.3} & $0.05-5.0$ & $0.6-5.5$ & $3.9-6.6$\\ [0.6ex]
       \multicolumn{1}{l}{Different-Vertex}  		& 1.5 	& 2.5^{+1.1}_{-0.7} & $0.04-11$ & $0.9-10$ & $2.5-7.4$\\[0.6ex]
       \multicolumn{1}{l}{Rest} 			& 11 	& \multicolumn{1}{c}{$15^{+6}_{-4}$} & $0.06-5.5$ & $1.1-22$ & \multicolumn{1}{c}{$14-27$}\\[0.6ex]
	\hline
       \hline
    \end{tabular}
        \label{tab:visiblexsection}
  \end{center}
\end{table*}

\subsection{Interpretations of the $Z^\prime_B$ and $Z'$-2HDM models}

\begin{figure}[!htbp]
  \centering
 \includegraphics[width=0.45\textwidth]{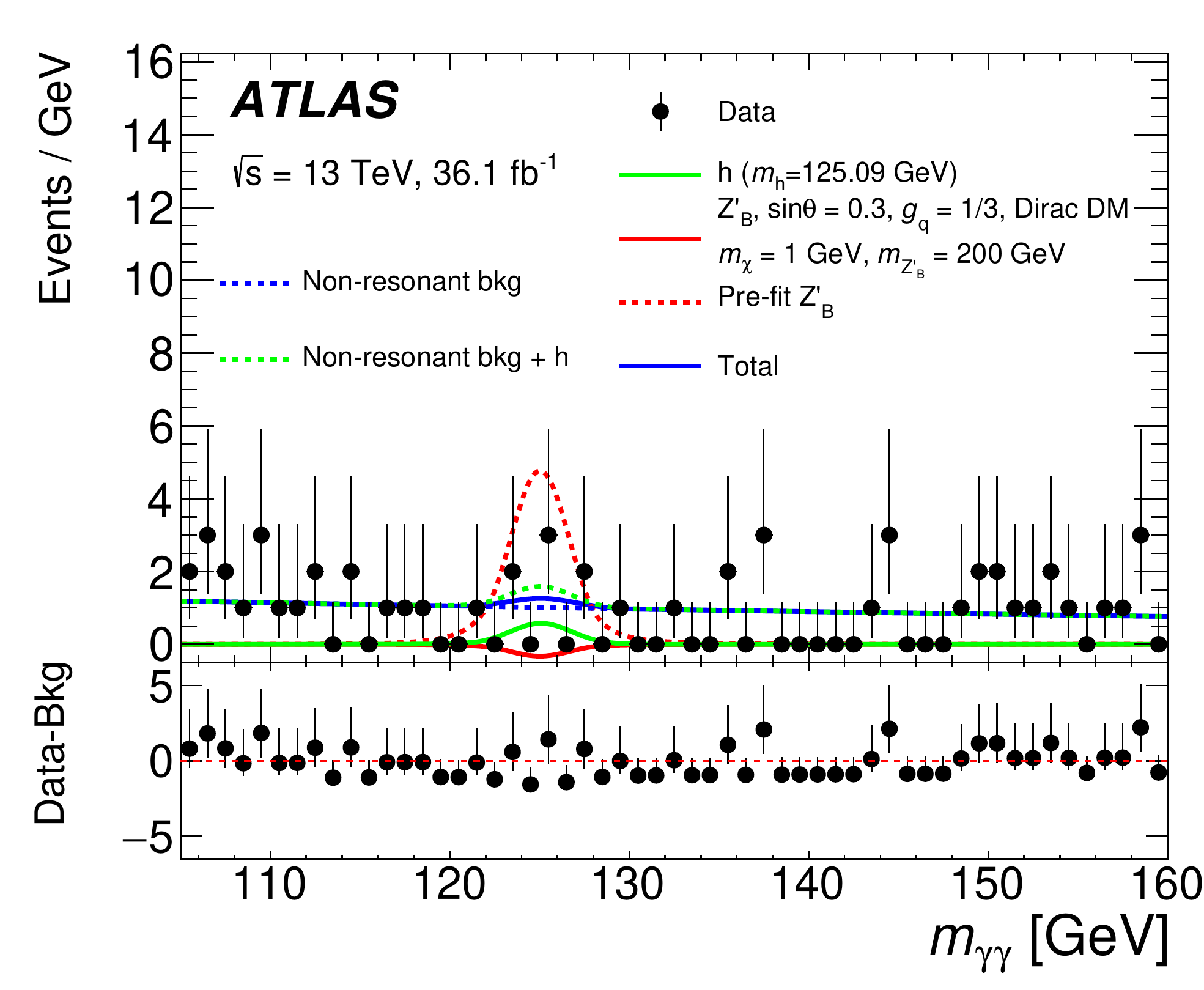}
 \caption{Diphoton invariant mass distribution for data and the corresponding fitted signal and background in
 the Mono-Higgs category for the $Z^\prime_B$ benchmark model fit using $g_q$ = 1/3, $g_{\chi}$ = 1, $\sin\theta$ = 0.3, and Dirac 
 fermion DM $m_\chi = 1$~\GeV\ as an illustration. A negative best-fit
 DM signal is found.
 The data is shown as dots with asymmetric error bars that represent central Poissonian confidence intervals at 68\% CL.
 The post-fitted signal (solid red line), pre-fitted signal (dashed red line), SM Higgs boson (solid green line),
 non-resonant background (dashed blue line) and the non-resonant background plus the SM Higgs boson (dashed green line)
 are shown as well as the total of all those contributions (solid blue line).
 In the bottom panel, the ``Bkg'' represents the total background including the SM Higgs boson productions.}
  \label{fig:res-data2}
\end{figure}

\begin{figure}[!htbp]
\centering
\begin{tabular}{cccc}
     	\subfigure[]{\includegraphics[width=0.49\textwidth]{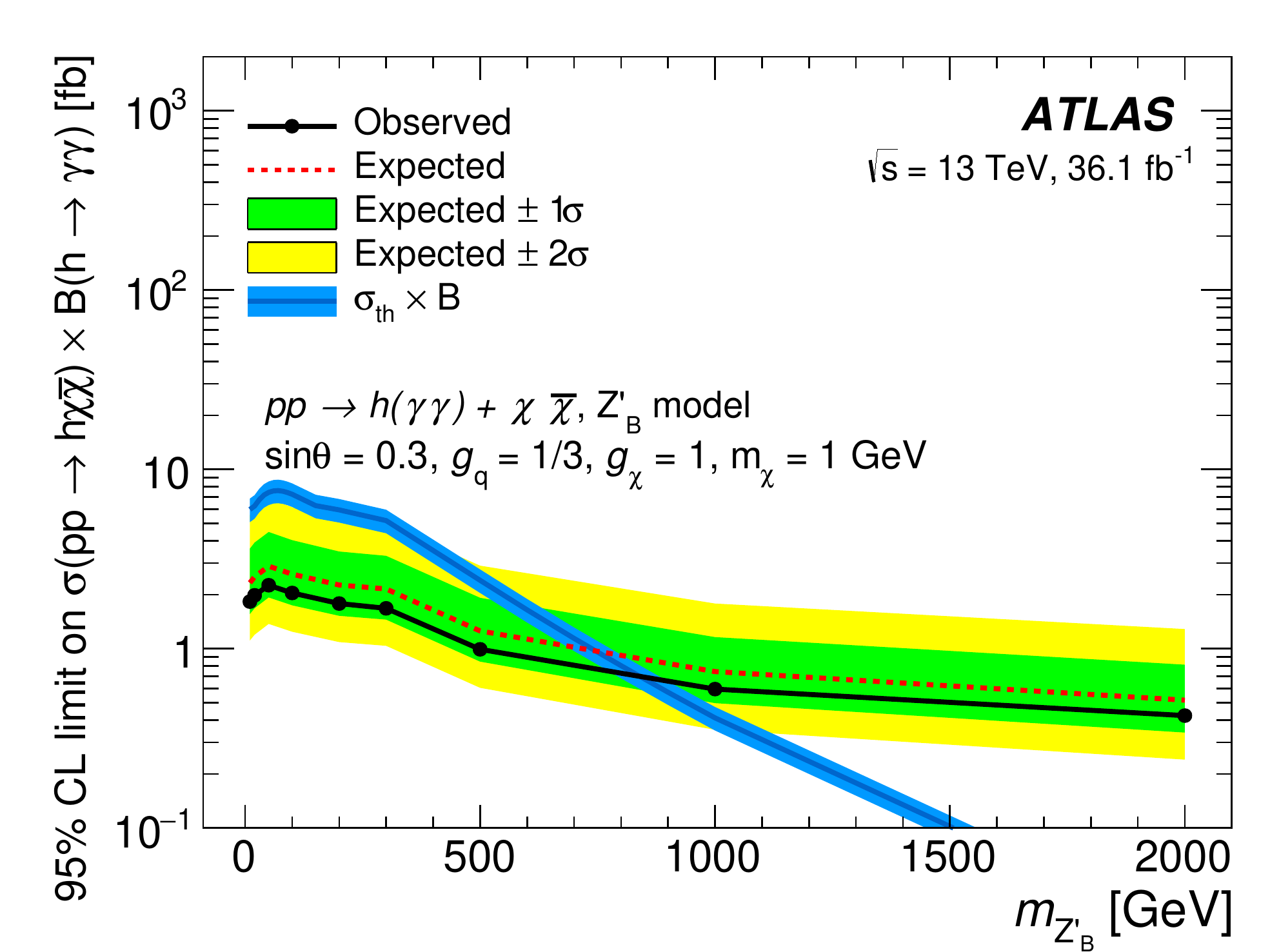}}
         \subfigure[]{\includegraphics[width=0.49\textwidth]{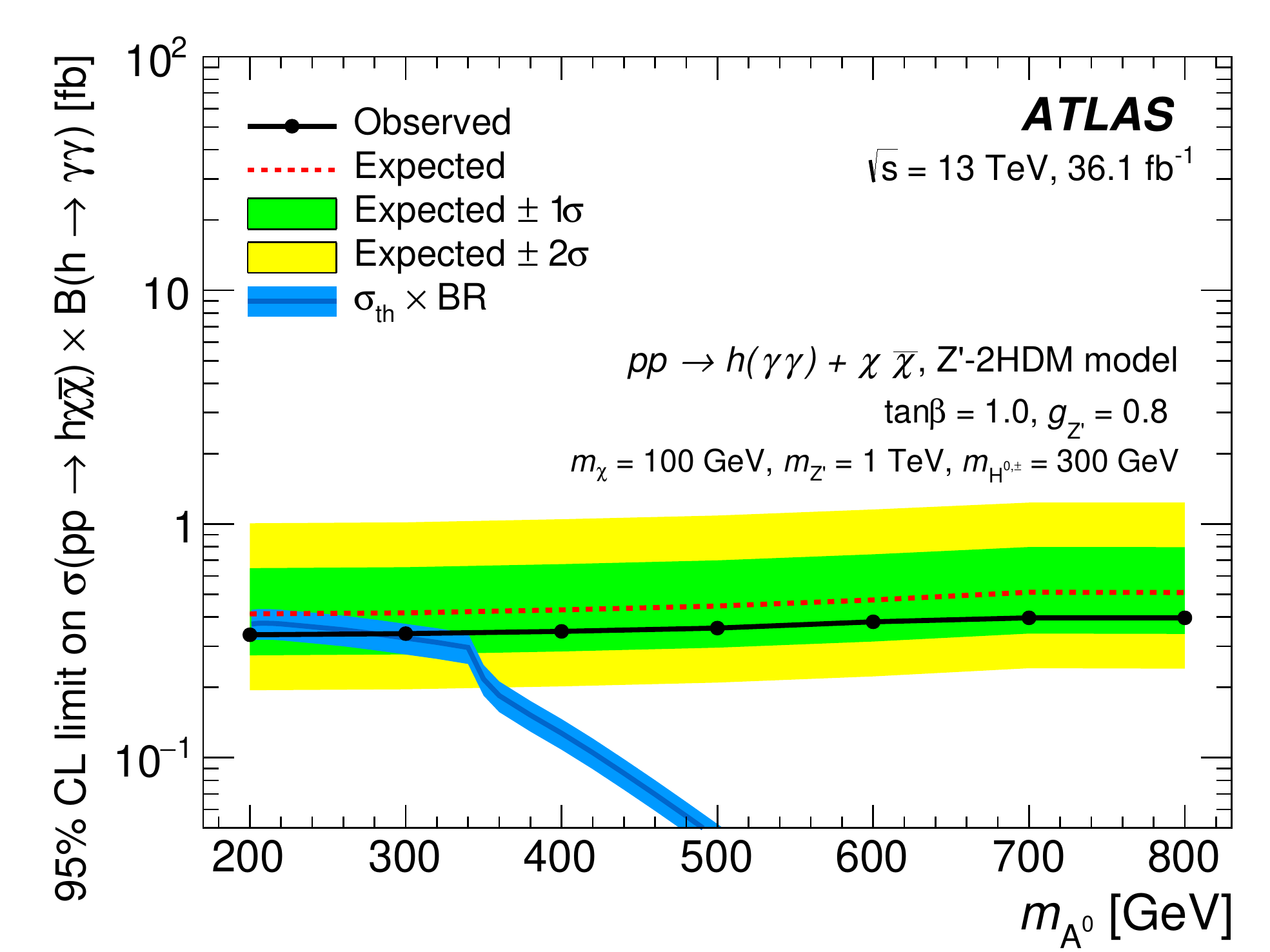}}
\end{tabular}
\caption{Expected (dashed lines) and observed (solid lines) $95\%$ CL upper limits on
	  $\sigma(pp \rightarrow h\chi\bar{\chi}) \times {\cal{B}} (h\rightarrow \gamma\gamma)$
	   for (a) the $Z'_B$ model for $g_q$ = 1/3, $g_{\chi}$ = 1,
           $\sin\theta$ = 0.3 and Dirac fermion DM $m_\chi = 1$~\GeV\;, and (b) the $Z'$-2HDM model
           for $\tan\beta$ = 1, $g_{Z'}$ = 0.8, $m_{Z'}$=1000~\GeV\ and Dirac
           fermion DM $m_\chi = 100$~\GeV, as a function of $m_{Z'_B}$ and $m_{A^0}$, respectively.
	   The masses of the neutral CP-even scalar ($H^0$) and the charged scalars ($H^{\pm}$) from $Z'$-2HDM model are set to 300 GeV.
	   The theoretical predictions of $\sigma(pp \rightarrow h\chi\bar{\chi}) \times {\cal{B}}(h\rightarrow \gamma\gamma)$
           for these two models (dark-blue lines with blue bands representing
           their associated theoretical systematic uncertainties) are also shown.
           The inset shows a zoomed-in view of the same figure in narrower ranges of both
           the $x$ and $y$ axes.
           }
   \label{fig:monoh_xsec_exclusion}
\end{figure}

Figure~\ref{fig:res-data2} shows the \mgg distributions in the Mono-Higgs category as well as the fits for a $Z^\prime_B$
benchmark point with $m_{Z'_{B}}$ = 200 GeV and $m_\chi$ = 1 GeV.
No significant excess is observed in this category. Upper limits are set on the
production cross sections in the two theoretical models considered.
Figure~\ref{fig:monoh_xsec_exclusion}(a) shows
the observed and median expected 95\% CL upper limits on \mbox{$\sigma(pp \rightarrow h\chi\bar{\chi}) \times {\cal{B}} (h\rightarrow \gamma\gamma)$}
as a function of the mediator mass $m_{Z'_{B}}$
for a DM mass of $1~\GeV$.
The cross sections times branching fraction of $h \rightarrow \gamma\gamma$ larger
than 2.3 fb are excluded for the full range of $m_{Z^\prime_B}$ between 10 and 2000 GeV at 95\% CL, and the $Z^\prime_B$ model is excluded
with $Z^\prime_B$ masses below 850~\GeV\ for a DM mass of 1~\GeV. 

In the $Z'$-2HDM scenario,
the observed and median expected 95\% CL upper limits on $\sigma(pp \rightarrow h\chi\bar{\chi}) \times {\cal{B}}(h\rightarrow \gamma\gamma)$
are shown in Fig.~\ref{fig:monoh_xsec_exclusion}(b),
as a function of the pseudoscalar boson mass $m_{A^0}$
for $m_{\chi}$ = $100~\GeV$ and $m_{Z'} = 1000~\GeV$.
The masses of the neutral CP-even scalar ($H^0$) and the charged scalars ($H^{\pm}$) from $Z'$-2HDM model are set to 300 GeV.
The theoretical cross section starts from $m_{A^0}=201~\GeV$.
The working point with $m_{A^0} = 200~\GeV$ is excluded since the resonant production of DM particles
at $m_{\chi}=100~\GeV$ significantly increases the cross section of the process.
To avoid the resonant regime where $m_{A^0} = 200~\GeV$ and $m_{\chi} = 100~\GeV$ and
allow a better limit interpolation, the point $m_{A^0} = 201~\GeV$ is shown in this plot instead of 200 GeV.
The drop of the theoretical prediction at $m_{A^0}=345$~\GeV\ is due
to a rapid change in the width when $A^0$ decaying to $t\bar{t}$ is kinematically allowed.
 Pseudoscalar boson masses below $280~\GeV$ are excluded for $m_{Z'} = 1000~\GeV$ and $m_{\chi}$ = $100~\GeV$.
Tables \ref{tab:limitsZprimeB} and \ref{tab:limits2hdm} present the 95\% CL observed and median expected limits
on $\sigma(pp \rightarrow h\chi\bar{\chi}) \times {\cal{B}}(h\rightarrow \gamma\gamma)$,
respectively, for the $Z^\prime_B$ benchmark model for different ${Z^\prime_B}$ masses
and the $Z^\prime$-2HDM model for different pseudoscalar ${A^0}$ masses.
The corresponding expected limits with one standard deviation are also shown.

\begin{figure}[!htbp]
\centering
\begin{tabular}{cccc}
\subfigure[]{\includegraphics[width=0.49\textwidth]{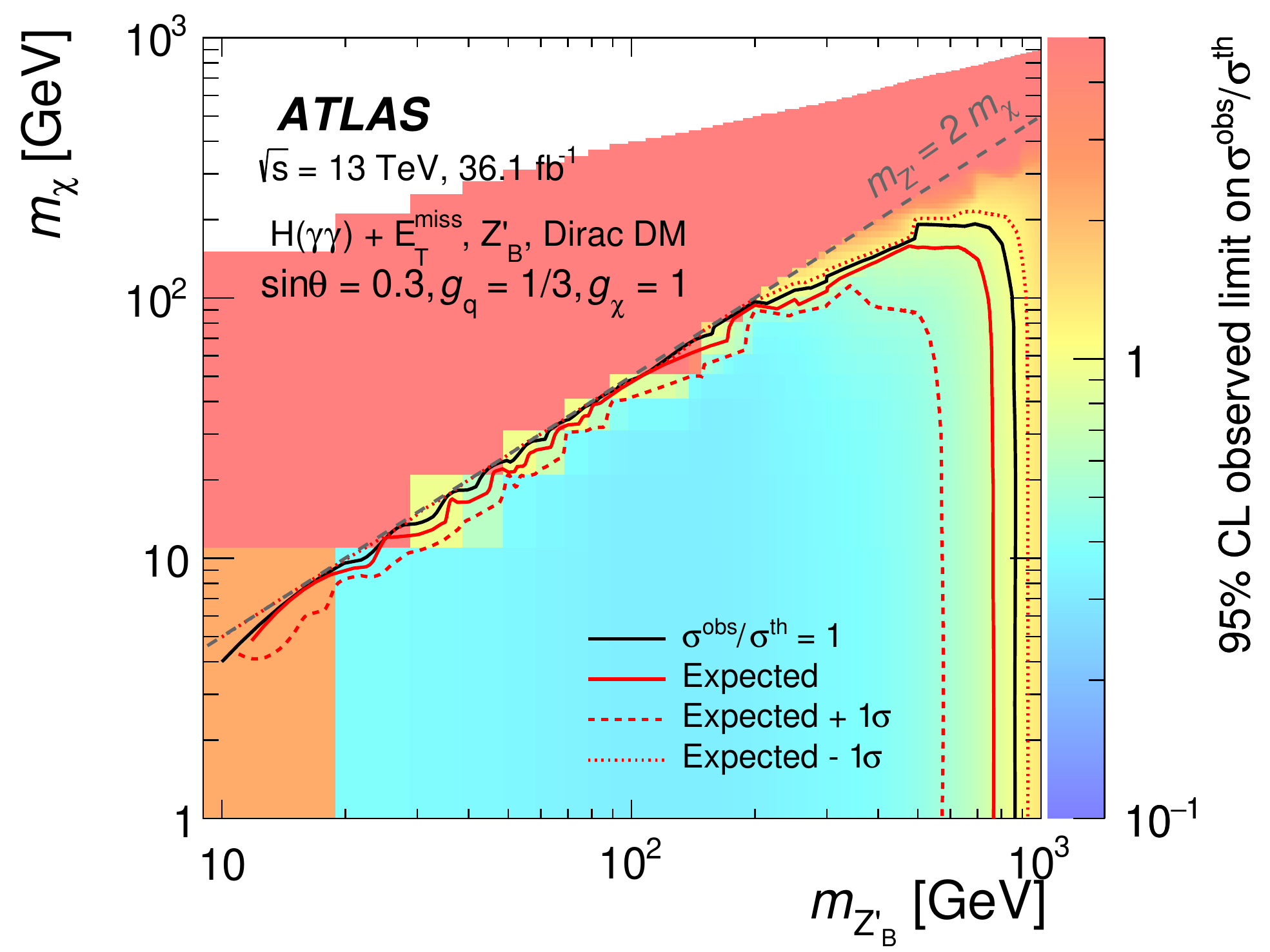}}
\subfigure[]{\includegraphics[width=0.49\textwidth]{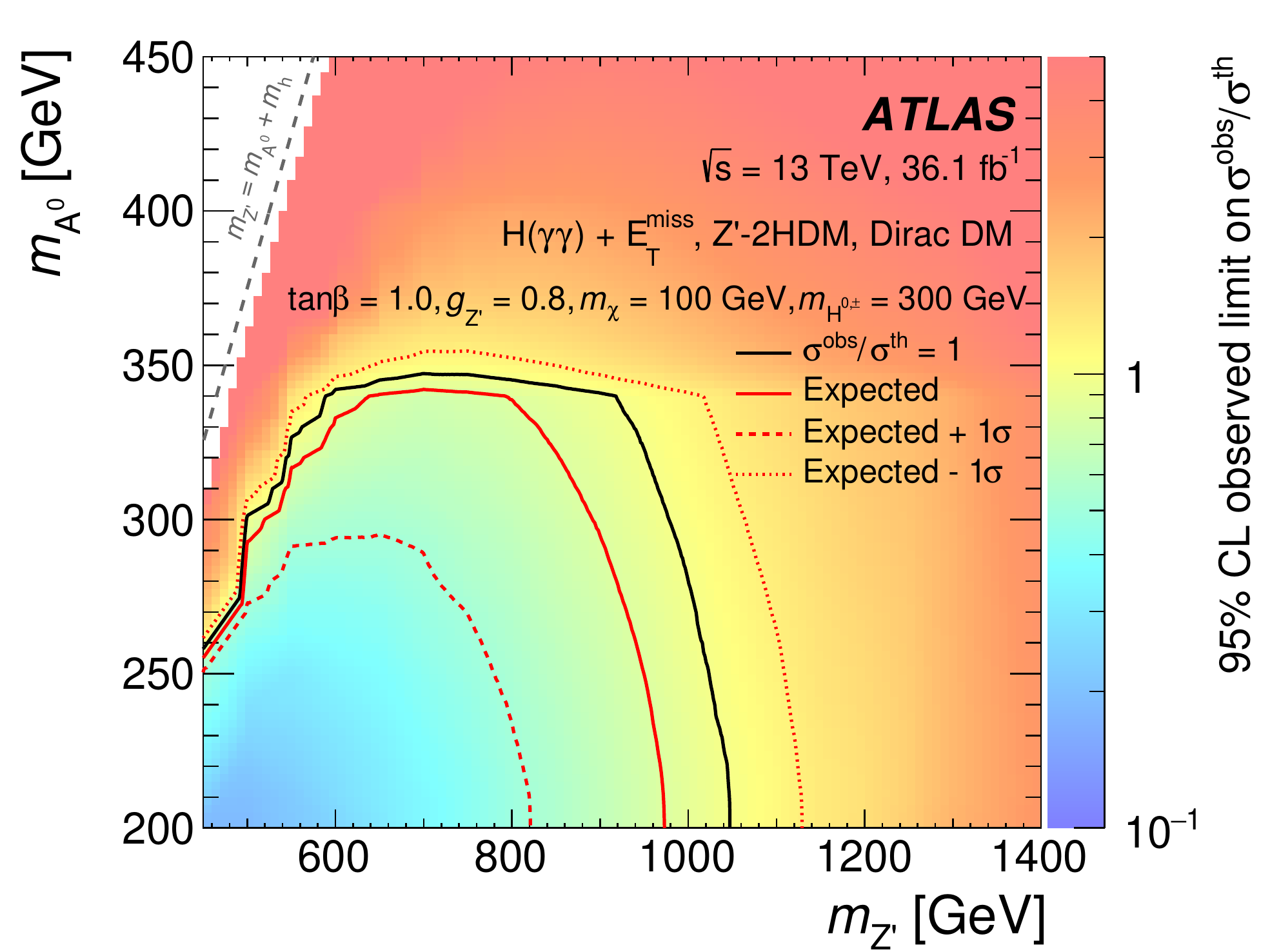}}
\end{tabular}
\caption{
The ratios of the observed and expected 95\% CL upper limits on the signal cross section
to the predicted signal cross sections for (a) the $Z'_{B}$ model in the ($m_\chi$, $m_{Z^{\prime}_{B}}$) plane
and (b) the $Z'$-2HDM model in the ($m_{A^0}$, $m_{Z'}$) plane.
For the $Z'_{B}$ model, the mixing angle $\sin\theta=0.3$, and the coupling values $g_q = 1/3$ and $g_{\chi}= 1$
are used. In the scenario of $Z'$-2HDM model, the ratio of the two-Higgs-doublet vacuum expectation values $\tan\beta = 1.0$,
Dirac fermion DM mass $m_{\chi}=100$ GeV, and the coupling value $g_{Z'} = 0.8$ are used.
The masses of the neutral CP-even scalar ($H^0$) and the charged scalars ($H^{\pm}$) from $Z'$-2HDM model are set to 300 GeV.
The plus and minus one standard deviation expected exclusion
curves are also shown as red dashed and dotted lines.
The regions below the lines (i.e.~with $\sigma^\text{obs}/\sigma^\text{th}<1$)
are excluded. In both figures, the gray dashed line corresponds to the boundary of the region above
which the $Z^\prime$ boson is produced off-shell.}
    \label{fig:monoh_xsec_exclusion_2D}
\end{figure}

A two-dimensional exclusion region in the plane formed by the mediator masses
and the DM particle mass is obtained by means of a reweighting
technique based on generator-level variables.
Samples for a variety of mass points are generated using the \mg generator matched to \pythiaeight using the A14
tune for parton showering and hadronization. Bin-by-bin weights for each of the different mass points are
obtained by comparing the generator level distribution of a given variable to the distribution for the same variable
in a fully simulated benchmark sample. This procedure can be repeated for other variables.
In the case of $Z'_{B}$ samples, weights are derived using the true \etmiss. In the case of $Z'$-2HDM samples,
weights are derived in a two-dimensional plane of the true \etmiss and \pTgg.
To validate this technique, several fully simulated and reconstructed samples
are produced and their reconstruction-level variables are compared with the sample obtained from the reweighting technique.
The acceptances of the samples after kinematic cuts agree within 5\%, and the residual difference is treated
as an extra systematic uncertainty in the signal yield.
The observed and expected 95\% CL limit contours for the signal strength
$\sigma^\text{obs}/\sigma^\text{th}$ are shown in
Fig.~\ref{fig:monoh_xsec_exclusion_2D} for both the $Z'_{B}$ and $Z'$-2HDM models, in which $\sigma^\text{obs}$ is the observed
limit on the model cross section at a given point of the parameter space and $\sigma^\text{th}$ is
the predicted cross section in the model at the same point.

\begin{table}[!htbp]
\centering
\caption{
Observed and expected 95\% CL upper limits (in fb) on
$\sigma(pp \rightarrow h\chi\bar{\chi}) \times {\cal{B}} (h\rightarrow \gamma\gamma)$
and associated expected $\pm 1\sigma$ upper limits for the $Z^\prime_B$ benchmark model for different $m_{Z^\prime_{B}}$
and for a fixed mass $m_{\chi}=1$~\GeV\ .}
\begin{tabular}{.....}
\toprule \toprule
\multicolumn{1}{l}{ $m_{Z'_{B}}$ [\GeV]} & \multicolumn{1}{l}{ Observed } & \multicolumn{1}{l}{ Expected } &
\multicolumn{1}{l}{ Expected +1$ \sigma$} & \multicolumn{1}{l}{ Expected $-1\sigma$ }\\
\midrule
10 & 1.83 & 2.33 & 3.61 & 1.56 \\
20 & 1.98 & 2.51 & 3.91 &  1.68 \\
50 & 2.26 & 2.88 & 4.47 &  1.93\\
100 & 2.04 & 2.60 & 4.03 & 1.74 \\
200 & 1.78 & 2.26 & 3.48 &  1.52 \\
300 & 1.67 & 2.15 & 3.29 & 1.45 \\
500 & 0.99 & 1.25 & 1.92 &  0.85 \\
1000 & 0.59 & 0.74 & 1.16 & 0.50 \\
2000 & 0.42 & 0.51 & 0.81 &  0.34 \\
\bottomrule \bottomrule
\end{tabular}
\label{tab:limitsZprimeB}
\end{table}

\begin{table}[!htbp]
\centering
\caption{
Observed and expected 95\% CL upper limits (in fb) on
$\sigma(pp \rightarrow h\chi\bar{\chi}) \times {\cal{B}} (h\rightarrow \gamma\gamma)$
and the associated expected $\pm 1\sigma$
upper limits for the $Z^\prime$-2HDM benchmark model for different $m_{A^0}$
and $m_{Z^{\prime}}=1000$~\GeV, $m_{\chi}=100$~\GeV.}
\begin{tabular}{.....}
\toprule \toprule
\multicolumn{1}{l}{ $m_{A^0}$ [\GeV]} & \multicolumn{1}{l}{ Observed } & \multicolumn{1}{l}{ Expected } & \multicolumn{1}{l}{ Expected +1$ \sigma$}
& \multicolumn{1}{l}{ Expected $-1\sigma$} \\
\midrule
200 &  0.33 & 0.41 & 0.65 & 0.27 \\
300 &  0.34 & 0.42 & 0.65 & 0.28 \\
400 &  0.35 & 0.43 & 0.67 & 0.28 \\
500 &  0.38 & 0.45 & 0.70 & 0.30 \\
600 &  0.39 & 0.47 & 0.74 & 0.31 \\
700 &  0.40 & 0.51 & 0.80 & 0.34 \\
800 &  0.40 & 0.51 & 0.80 & 0.34 \\
\bottomrule \bottomrule
\end{tabular}
\label{tab:limits2hdm}
\end{table}

\begin{figure}[!htbp]
\centering
\begin{tabular}{cccc}
\includegraphics[width=0.49\textwidth]{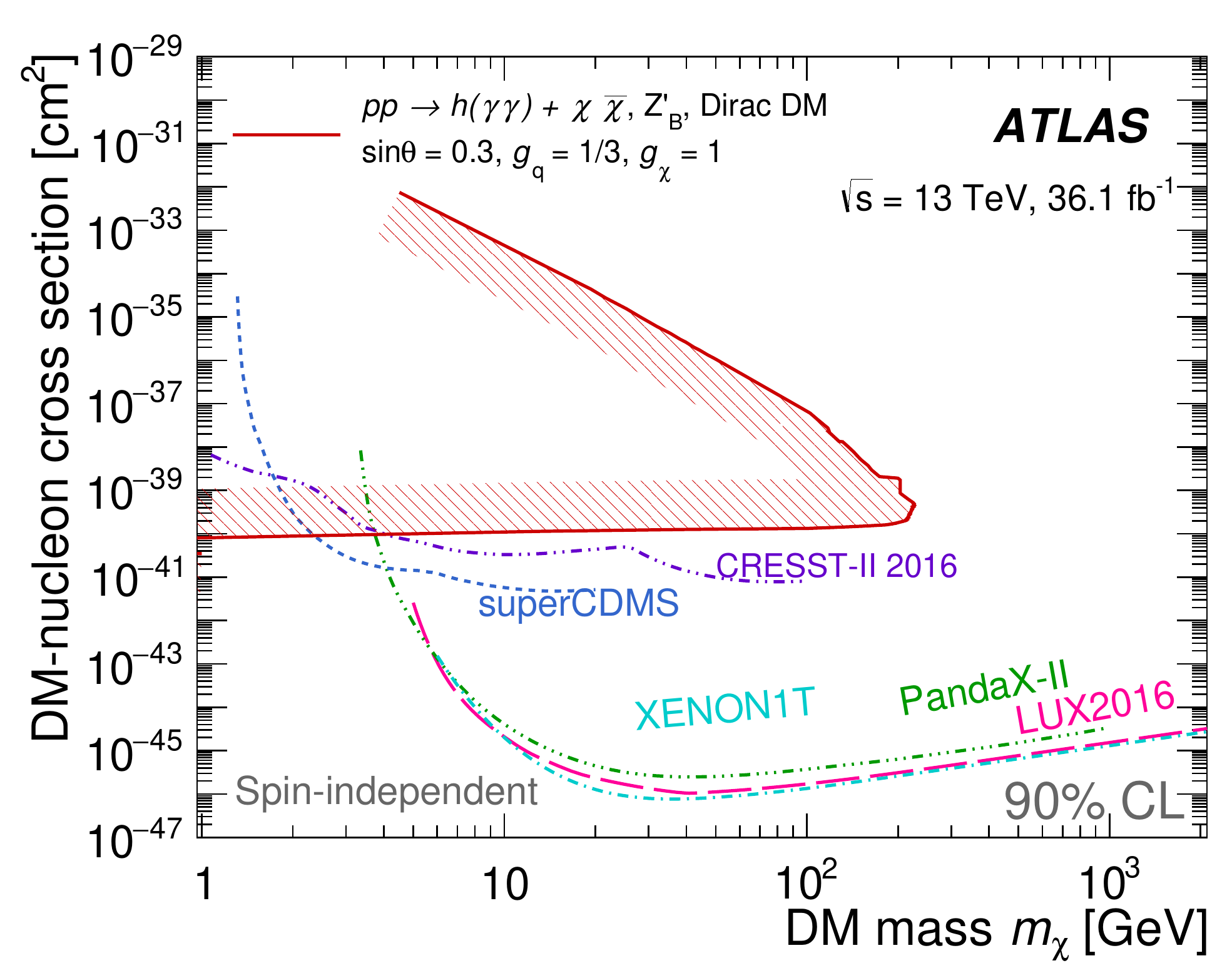}
\end{tabular}
 \caption{
 A comparison of the inferred limits to the constraints from direct detection
 experiments on the spin-independent DM--nucleon cross
section in the context of the $Z'_B$ simplified model with vector couplings. Limits are shown at 90\% CL.
The results from this analysis, in which the region inside the contour is excluded, are compared with limits from the LUX~\cite{Akerib:2016vxi}, PandaX-II~\cite{Tan:2016zwf},
XENON~\cite{Aprile:2017iyp}, superCDMS~\cite{Agnese:2015nto}, and CRESST-II~\cite{Angloher:2015ewa} experiments.
The comparison is model-dependent and solely valid in the context of this model,
assuming Dirac fermion DM, mixing angle $\sin\theta=0.3$, and the coupling values $g_q = 1/3$ and $g_{\chi}= 1$.
The impact of renormalization-group evolution effects~\cite{Crivellin:2014qxa,DEramo:2014nmf} when comparing
collider and direct detection limits is not taken into consideration here.
}
    \label{fig:monoh_DM_NucleonXS_SI}
\end{figure}

Figure~\ref{fig:monoh_DM_NucleonXS_SI} shows a comparison of the inferred limits to the
constraints from direct detection experiments on the spin-independent (SI) DM--nucleon cross
section in the context of the $Z'_B$ simplified model with vector couplings using the relation~\cite{monoh}:

\begin{equation*}
\sigma^\text{SI}_{N\chi} = \frac{\mu^{2}_{N\chi}}{\pi A^2}\left[Zf_{p}-(A-Z)f_{n}\right]^2 \,,
\end{equation*}

in which $\mu_{N\chi}=m_{\chi}m_{N}/(m_{\chi}+m_{N})$ is the reduced mass of the
DM--nucleon system, and ${f_{p,n}=3g_{q}g_{\chi}/m^{2}_{Z'_B}}$ are the couplings
between DM particles and protons and neutrons.
The parameter $Z$ is the number of protons in the considered nucleus and $A$ the number of nucleons (both set to 1).
Limits are shown at 90\% CL.
For comparison, results from direct detection experiments (LUX~\cite{Akerib:2016vxi}, PandaX-II~\cite{Tan:2016zwf},
XENON~\cite{Aprile:2017iyp}, superCDMS~\cite{Agnese:2015nto}, and CRESST-II~\cite{Angloher:2015ewa}) are also shown.
The comparison is model-dependent and solely valid in the context of this model.
The results for the $Z'_B$ model, with couplings $g_{q} = 1/3$ and $g_{\chi}=1$ for this search,
are more stringent than direct detection experiments for $m_{\chi} < 2.5$~GeV
and extend to DM masses well below 1 GeV.
The shape of the exclusion line at DM mass $m_{\chi} \sim 200~\GeV$ to low
masses is due to the loss of sensitivity in $Z'_B$ models where DM particles are produced via an off-shell process.
The impact of renormalization-group evolution effects~\cite{Crivellin:2014qxa,DEramo:2014nmf} when comparing
collider and direct detection limits is not taken into consideration here.

\begin{figure}[!htbp]
  \centering
     \begin{tabular}{cccc}
     \subfigure[]{\includegraphics[width=0.45\textwidth]{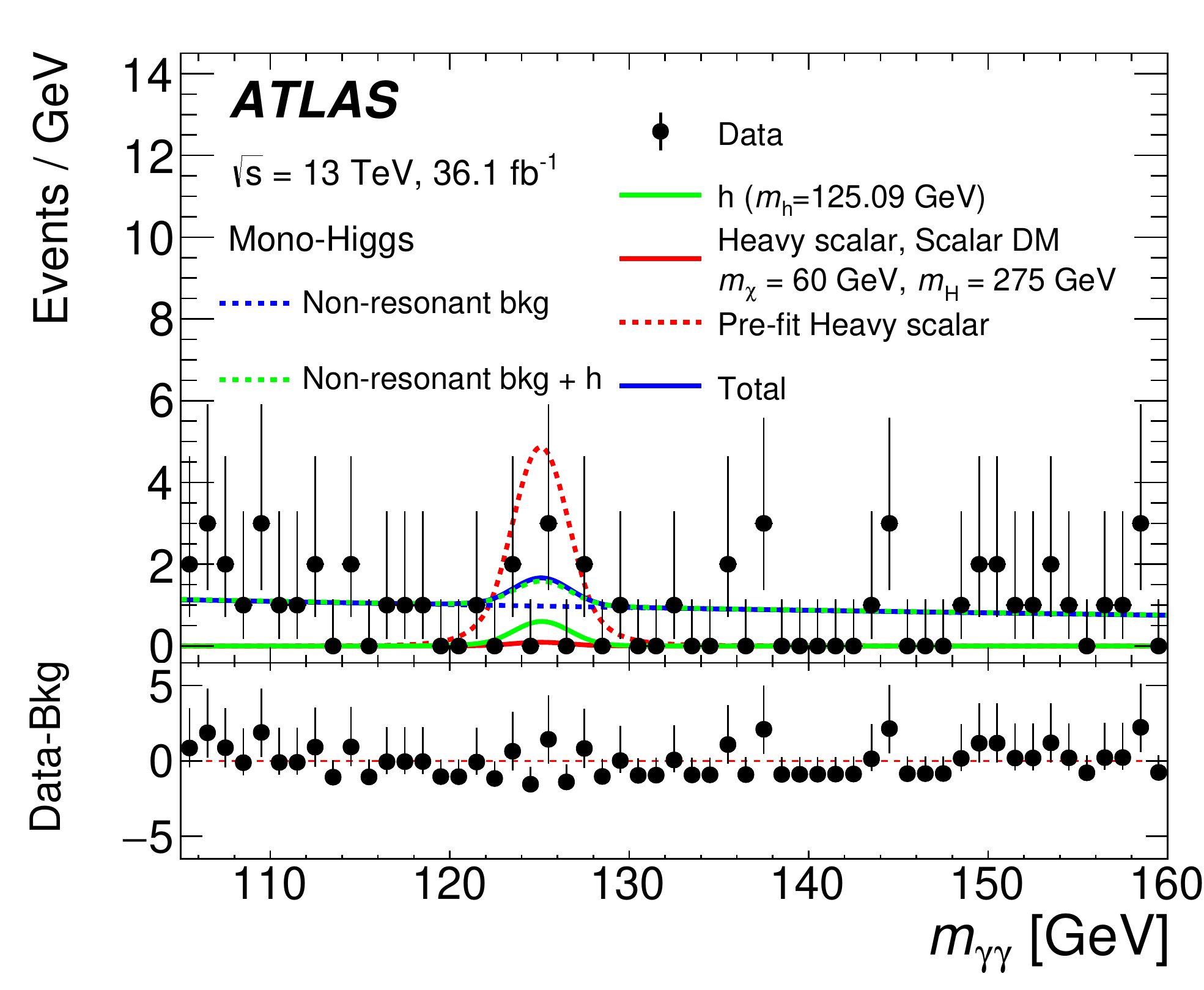}} \\
     \subfigure[]{\includegraphics[width=0.45\textwidth]{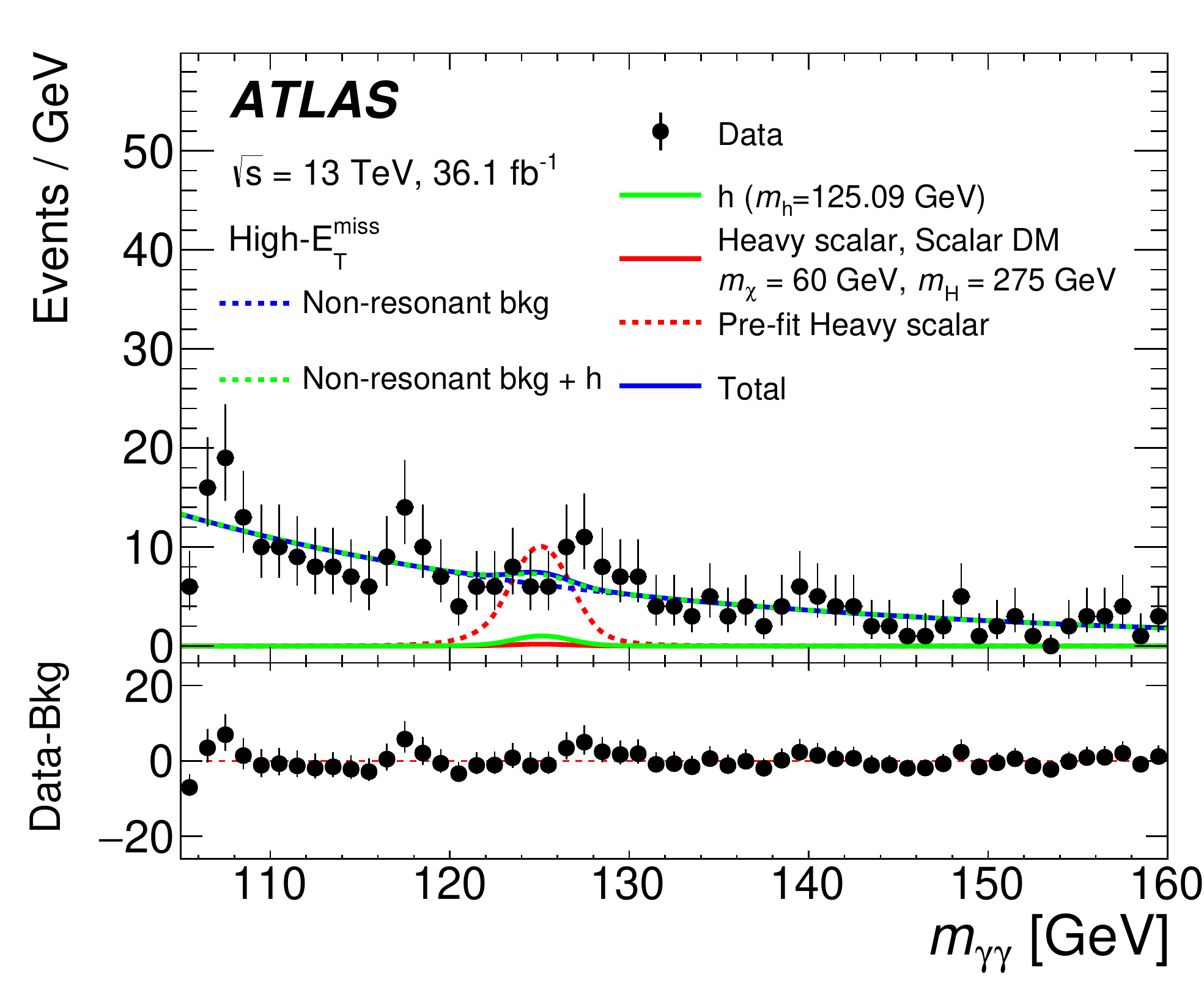}}
     \subfigure[]{\includegraphics[width=0.45\textwidth]{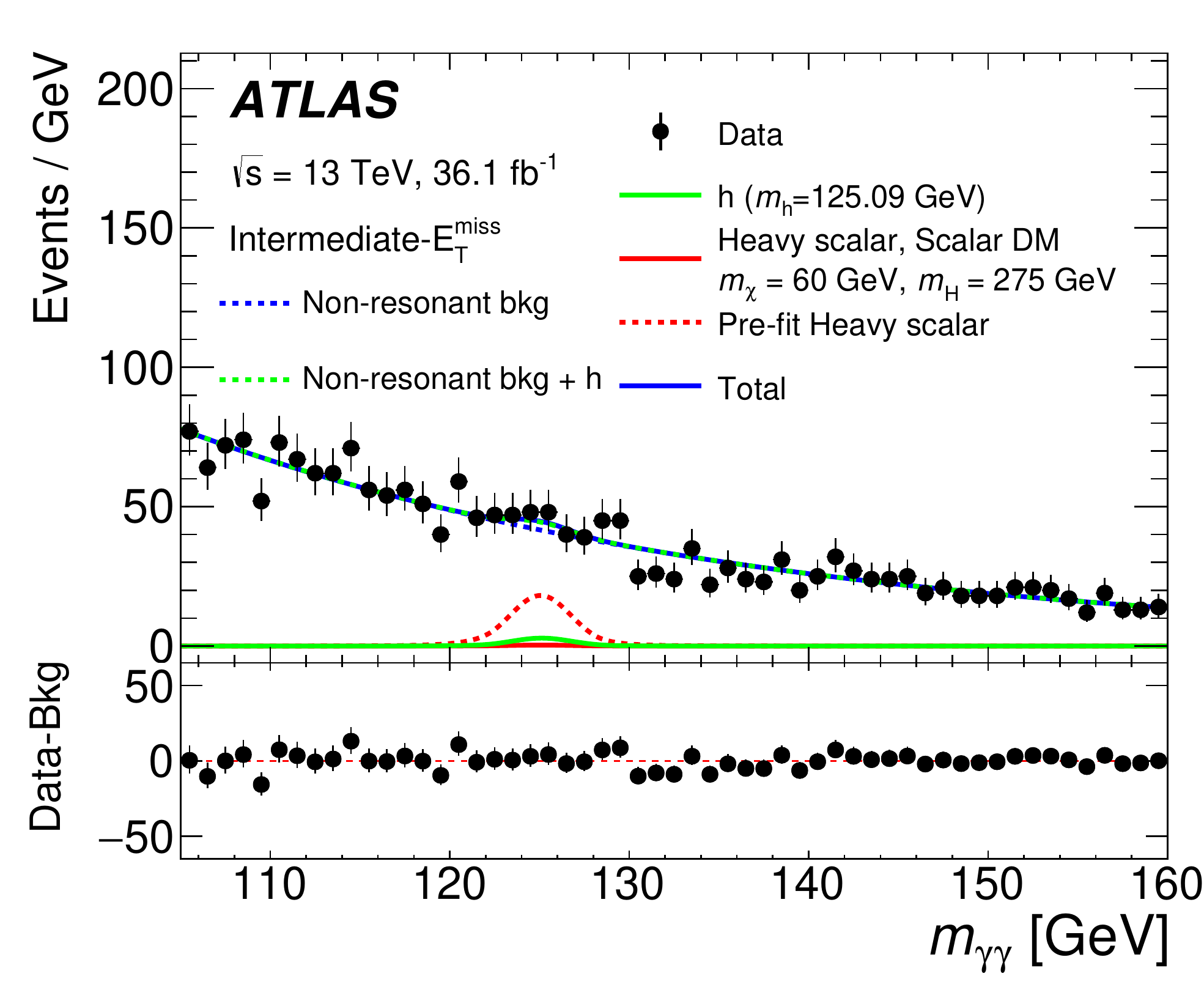}} \\
     \subfigure[]{\includegraphics[width=0.45\textwidth]{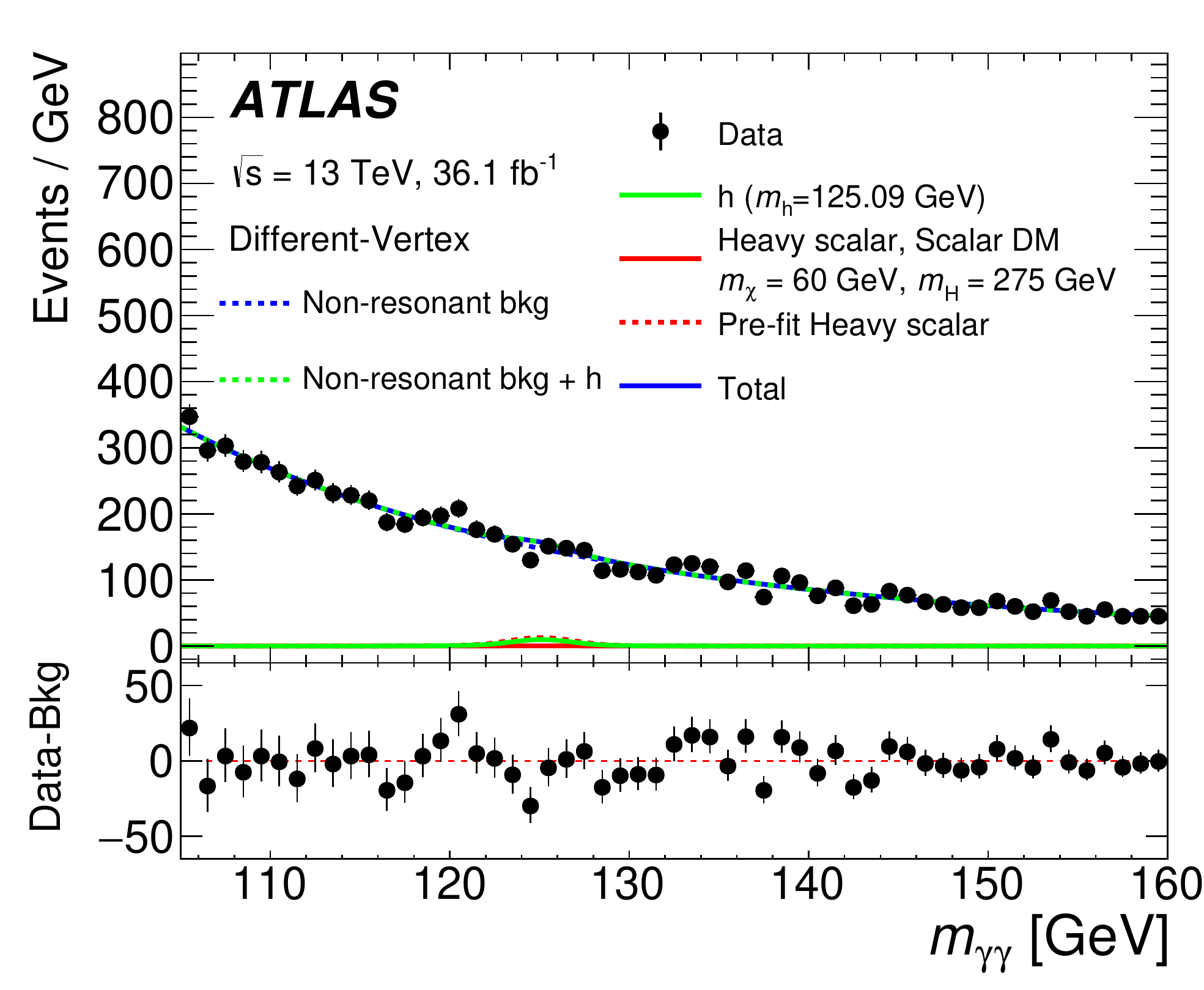}}
     \subfigure[]{\includegraphics[width=0.45\textwidth]{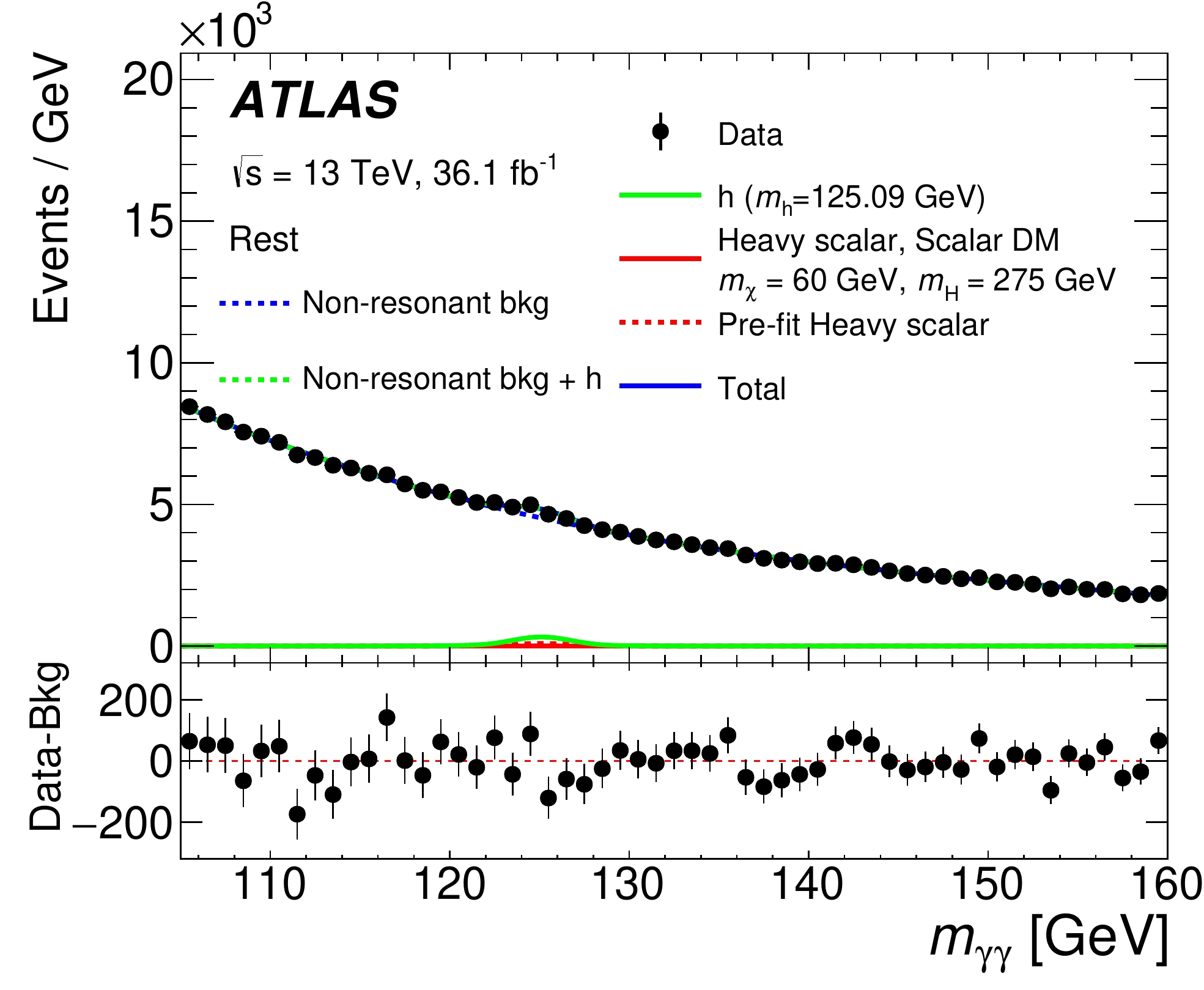}}
     \end{tabular}
     \caption{Diphoton invariant mass distribution for data and the corresponding fitted signal and background in the
           five categories, (a) Mono-Higgs category, (b) High-\etmiss category, (c) Intermediate-\etmiss category,
           (d) Different-Vertex category, and (e) Rest category.
           In each plot, the data (dots with asymmetric error bars) is shown.
           The error bars represent the central Poissonian confidence intervals at
            68\% CL. The simultaneous fit result including a heavy-scalar signal (solid red line),
	   SM Higgs boson (solid green line), the non-resonant background (dashed blue line), and the non-resonant
           background plus the SM Higgs boson (dashed green line) are shown
           as well as the sum of all those contributions (solid blue line).
	   In the bottom panel, the ``Bkg'' represents the total background including the SM Higgs boson productions.
           The both pre- and post-fitted heavy-scalar signals shown here correspond to $\mBH = 275~$\GeV\  and scalar DM $m_\chi = 60~$\GeV.
        } 
  \label{fig:res-data}
\end{figure}

\begin{figure}[!htbp]
  \centering
    \begin{tabular}{cccc}
    \includegraphics[width=0.49\textwidth]{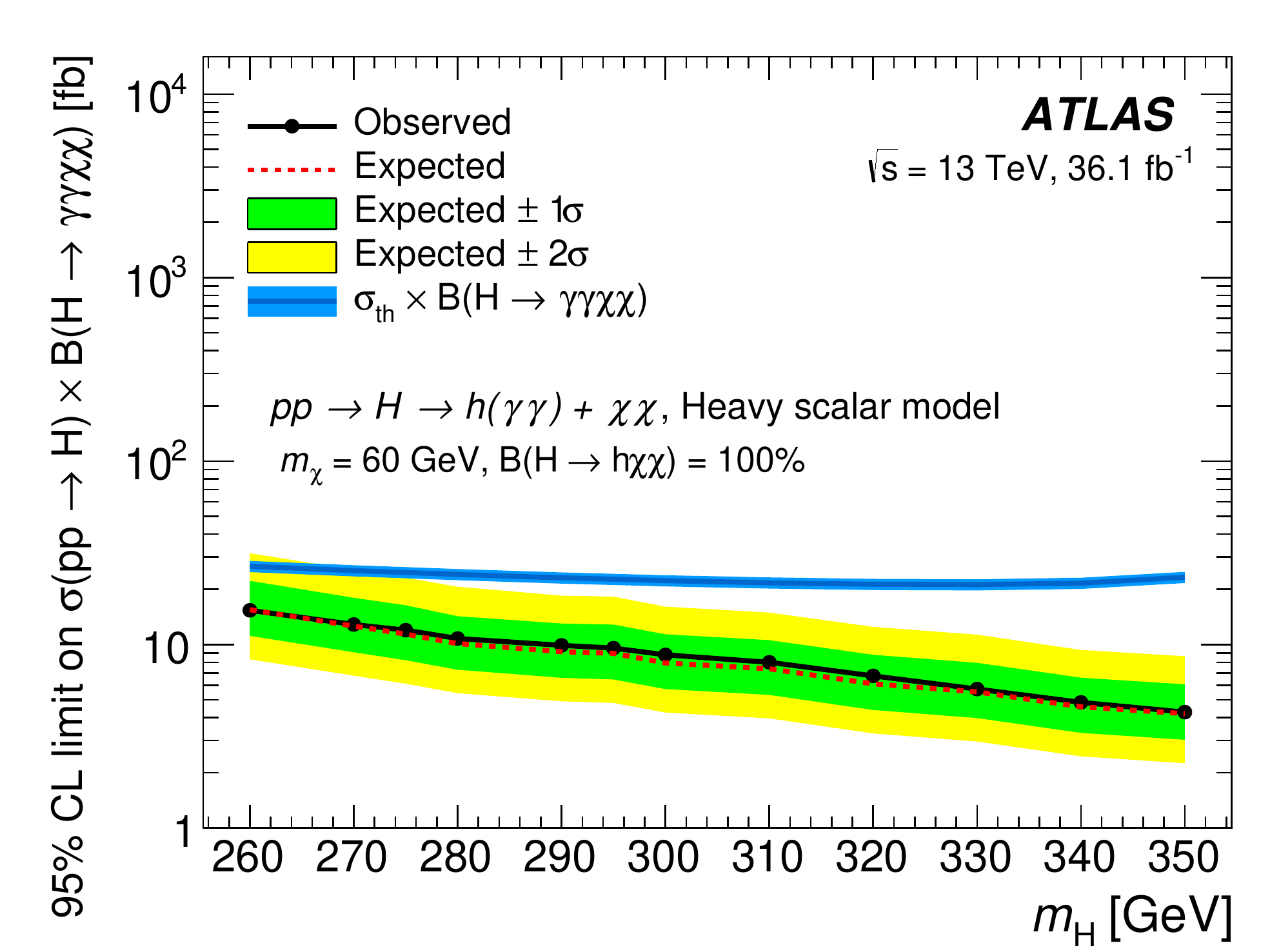}
    \end{tabular}
  \caption{
          Observed and expected 95\% CL upper limits on
          the $\sigma(pp \rightarrow H) \times {\cal{B}} (H \rightarrow \gamma \gamma \chi\chi)$
	   with a scalar DM candidate mass of $60$~\GeV\ as a function of the
           heavy-scalar-boson mass in the range 260 $< \mBH <$ 350 \GeV.
	   A 100\% branching fraction is assumed for $H \rightarrow h\chi\chi$.
            The theoretical prediction for the model (dark-blue lines with blue bands representing
           their associated theoretical systematic uncertainties) is also shown. The
           theoretical cross section is assumed to be equal to that of a SM Higgs boson with the same mass
           produced in gluon--gluon fusion.
          }
  \label{fig:res-xsex-excl}
\end{figure}

\subsection{Interpretation of the heavy-scalar model}

Figure~\ref{fig:res-data} shows the \mgg distributions in the five categories as
well as the fitted contribution of a heavy-scalar boson for illustration.
No significant excess is observed in any category.
In the heavy-scalar interpretation,
the 95\% CL upper limits on the $\sigma(pp \rightarrow H) \times {\cal{B}} (H \rightarrow \gamma \gamma \chi\chi)$
as a function of \mBH for $\mchi$ = 60 \GeV\ are shown in Fig.~\ref{fig:res-xsex-excl} and Table~\ref{tab:limitsHS60}, where a 100\% branching fraction is assumed for $H \rightarrow h\chi\chi$.
The upper limit at 95\% CL is $15.4$~fb for $\mBH=260$~\GeV, and $4.3$~fb for
$\mBH=350$~\GeV.

\begin{table}[h!tb]
\centering
\caption{
Observed and expected 95\% CL upper limits (in fb) on the $\sigma(pp \rightarrow H) \times {\cal{B}} (H \rightarrow \gamma \gamma \chi\chi)$,
where a 100\% branching fraction is assumed for $H \rightarrow
h\chi\chi$, and the associated expected $\pm 1\sigma$ upper limits for
the heavy-scalar model for different mediator masses and $m_{\chi}=60$~\GeV.}
\begin{tabular}{.....}
\toprule \toprule
\multicolumn{1}{l}{ $\mBH$ [\GeV]} & \multicolumn{1}{l}{ Observed} & \multicolumn{1}{l}{Expected} & \multicolumn{1}{l}{Expected +1$ \sigma$} & \multicolumn{1}{l}{Expected $-1\sigma$}\\
\midrule
260 & 15.4 & 15.5  &  22.3  &  11.2 \\
270 & 12.9 & 12.6  & 18.0   &   9.1  \\
275 & 12.0 & 11.4  & 16.4   &    8.2 \\
285 & 10.8 & 11.1  & 14.3   &    7.3 \\
290 &   9.9 &   9.1  & 13.0   &    6.6 \\
295 &   9.6 &   9.0  & 12.8   &    6.5 \\
300 &   8.8 &   7.9  & 11.4   &    5.7 \\
310 &   8.0 &   7.4  & 10.6   &    5.3 \\
320 &   6.8 &   6.1  &   8.8   &    4.4 \\
330 &   5.7 &   5.5  &   7.9   &    4.0 \\
340 &   4.8 &   4.6  &   6.6   &    3.3 \\
350 &   4.3 &   4.2  &   6.1   &    3.0 \\
\bottomrule \bottomrule
\end{tabular}
\label{tab:limitsHS60}
\end{table}
 \clearpage

\section{Summary}
\label{sec:summary}

A search for dark matter in association with a Higgs boson decaying to two photons is presented.
This study is based on data collected with the ATLAS detector, corresponding to an integrated luminosity of $36.1~\mathrm{fb}^{-1}$ of proton--proton
collisions at the LHC at a center-of-mass energy of 13 TeV in 2015 and 2016.
No significant excess over the expected background is observed.
For the Mono-Higgs category, a visible cross section larger than 0.19 fb is
excluded at 95\% CL for BSM physics processes producing missing transverse
momentum and a SM Higgs boson decaying into two photons.
Upper limits at 95\% CL are also set on
the production cross section times branching fraction
of the Higgs boson decaying into two photons in association with missing
transverse momentum in three different benchmark models:
a $Z^\prime_B$ model, a $Z^\prime$-2HDM model and a heavy scalar boson ($H$) model.
Limits at 95\% CL are also set on the observed signal strength in a two-dimensional
$m_\chi$--$m_{Z^\prime_B}$ plane for the $Z^\prime_B$ model, and the
$m_{A^0}$--$m_{Z^\prime}$ plane for the $Z^\prime$-2HDM model.
Additionally, the results for the $Z^\prime_B$ model are interpreted in terms
of 90\% CL limits on
the dark-matter--nucleon scattering cross section, as a function of the dark-matter-particle mass, for a spin-independent scenario.
For a dark-matter mass lower than 2.5~\GeV, the constraint with couplings $g_{q} = 1/3$ and $g_{\chi}=1$
placed on the DM--nucleon cross section is more stringent than limits from direct detection experiments.
In the model involving the production of a heavy scalar boson, 95\% CL upper
limits are set on the production cross section times the branching fraction of
$H \rightarrow h\chi\chi \rightarrow \gamma \gamma \chi\chi$
for a dark-matter particle with mass of
60~\GeV, where a 100\% branching fraction is assumed for $H \rightarrow h\chi\chi$.
The heavy-scalar model assuming $H$ production through gluon--gluon fusion with
a cross section identical to that of a SM Higgs boson of the same mass, is excluded for all the benchmark points
investigated.
 
\section*{Acknowledgements}

We thank CERN for the very successful operation of the LHC, as well as the
support staff from our institutions without whom ATLAS could not be
operated efficiently.

We acknowledge the support of ANPCyT, Argentina; YerPhI, Armenia; ARC, Australia; BMWFW and FWF, Austria; ANAS, Azerbaijan; SSTC, Belarus; CNPq and FAPESP, Brazil; NSERC, NRC and CFI, Canada; CERN; CONICYT, Chile; CAS, MOST and NSFC, China; COLCIENCIAS, Colombia; MSMT CR, MPO CR and VSC CR, Czech Republic; DNRF and DNSRC, Denmark; IN2P3-CNRS, CEA-DSM/IRFU, France; SRNSF, Georgia; BMBF, HGF, and MPG, Germany; GSRT, Greece; RGC, Hong Kong SAR, China; ISF, I-CORE and Benoziyo Center, Israel; INFN, Italy; MEXT and JSPS, Japan; CNRST, Morocco; NWO, Netherlands; RCN, Norway; MNiSW and NCN, Poland; FCT, Portugal; MNE/IFA, Romania; MES of Russia and NRC KI, Russian Federation; JINR; MESTD, Serbia; MSSR, Slovakia; ARRS and MIZ\v{S}, Slovenia; DST/NRF, South Africa; MINECO, Spain; SRC and Wallenberg Foundation, Sweden; SERI, SNSF and Cantons of Bern and Geneva, Switzerland; MOST, Taiwan; TAEK, Turkey; STFC, United Kingdom; DOE and NSF, United States of America. In addition, individual groups and members have received support from BCKDF, the Canada Council, CANARIE, CRC, Compute Canada, FQRNT, and the Ontario Innovation Trust, Canada; EPLANET, ERC, ERDF, FP7, Horizon 2020 and Marie Sk{\l}odowska-Curie Actions, European Union; Investissements d'Avenir Labex and Idex, ANR, R{\'e}gion Auvergne and Fondation Partager le Savoir, France; DFG and AvH Foundation, Germany; Herakleitos, Thales and Aristeia programmes co-financed by EU-ESF and the Greek NSRF; BSF, GIF and Minerva, Israel; BRF, Norway; CERCA Programme Generalitat de Catalunya, Generalitat Valenciana, Spain; the Royal Society and Leverhulme Trust, United Kingdom.

The crucial computing support from all WLCG partners is acknowledged gratefully, in particular from CERN, the ATLAS Tier-1 facilities at TRIUMF (Canada), NDGF (Denmark, Norway, Sweden), CC-IN2P3 (France), KIT/GridKA (Germany), INFN-CNAF (Italy), NL-T1 (Netherlands), PIC (Spain), ASGC (Taiwan), RAL (UK) and BNL (USA), the Tier-2 facilities worldwide and large non-WLCG resource providers. Major contributors of computing resources are listed in Ref.~\cite{ATL-GEN-PUB-2016-002}.

\printbibliography

\newpage \begin{flushleft}
{\Large The ATLAS Collaboration}

\bigskip

M.~Aaboud$^\textrm{\scriptsize 137d}$,
G.~Aad$^\textrm{\scriptsize 88}$,
B.~Abbott$^\textrm{\scriptsize 115}$,
O.~Abdinov$^\textrm{\scriptsize 12}$$^{,*}$,
B.~Abeloos$^\textrm{\scriptsize 119}$,
S.H.~Abidi$^\textrm{\scriptsize 161}$,
O.S.~AbouZeid$^\textrm{\scriptsize 139}$,
N.L.~Abraham$^\textrm{\scriptsize 151}$,
H.~Abramowicz$^\textrm{\scriptsize 155}$,
H.~Abreu$^\textrm{\scriptsize 154}$
% ...
\bigskip
\\
% ...
$^{12}$ Institute of Physics, Azerbaijan Academy of Sciences, Baku, Azerbaijan\\
% ...
$^{88}$ CPPM, Aix-Marseille Universit{\'e} and CNRS/IN2P3, Marseille, France\\
% ...
$^{115}$ Homer L. Dodge Department of Physics and Astronomy, University of Oklahoma, Norman OK, United States of America\\
% ...
$^{119}$ LAL, Univ. Paris-Sud, CNRS/IN2P3, Universit{\'e} Paris-Saclay, Orsay, France\\
% ...
$^{137}$ $^{(a)}$ Facult{\'e} des Sciences Ain Chock, R{\'e}seau Universitaire de Physique des Hautes Energies - Universit{\'e} Hassan II, Casablanca; $^{(b)}$ Centre National de l'Energie des Sciences Techniques Nucleaires, Rabat; $^{(c)}$ Facult{\'e} des Sciences Semlalia, Universit{\'e} Cadi Ayyad, LPHEA-Marrakech; $^{(d)}$ Facult{\'e} des Sciences, Universit{\'e} Mohamed Premier and LPTPM, Oujda; $^{(e)}$ Facult{\'e} des sciences, Universit{\'e} Mohammed V, Rabat, Morocco\\
% ...
$^{139}$ Santa Cruz Institute for Particle Physics, University of California Santa Cruz, Santa Cruz CA, United States of America\\
% ...
$^{151}$ Department of Physics and Astronomy, University of Sussex, Brighton, United Kingdom\\
% ...
$^{154}$ Department of Physics, Technion: Israel Institute of Technology, Haifa, Israel\\
$^{155}$ Raymond and Beverly Sackler School of Physics and Astronomy, Tel Aviv University, Tel Aviv, Israel\\
% ...
$^{161}$ Department of Physics, University of Toronto, Toronto ON, Canada\\
% ...
$^{a}$ Also at Department of Physics, King's College London, London, United Kingdom\\
$^{b}$ Also at Institute of Physics, Azerbaijan Academy of Sciences, Baku, Azerbaijan\\
$^{c}$ Also at Novosibirsk State University, Novosibirsk, Russia\\
$^{d}$ Also at TRIUMF, Vancouver BC, Canada\\
$^{e}$ Also at Department of Physics {\&} Astronomy, University of Louisville, Louisville, KY, United States of America\\
% ...
$^{*}$ Deceased
\end{flushleft}

 \end{document}